\documentclass[a4paper,12pt]{article}
\usepackage{graphicx}
\usepackage{amsmath}
\usepackage{amssymb}
\usepackage{array}
\usepackage{euscript}
\usepackage{subfigure}
\usepackage{pifont}
\usepackage{times}
\usepackage{geometry}
\newcommand{\rmM}{\text{\fontfamily{cmr}\selectfont M}}
\newcommand{\rmi}{\text{\fontfamily{cmr}\selectfont i}}
\newcommand{\rme}{\text{\fontfamily{cmr}\selectfont e}}
\newcommand{\rmd}{\text{\fontfamily{cmr}\selectfont d}}
\newcommand{\sgn}{\text{\fontfamily{cmr}\selectfont sgn}}

\newcommand{\arc}{\text{\fontfamily{cmr}\selectfont arc\!}}
\newcommand{\Eul}[1]{\EuScript{#1}}
\newcommand{\modu}{\text{\fontfamily{cmr}\selectfont \,mod\,}}
\newcommand{\RE}[1]{\text{\fontfamily{cmr}\selectfont Re}\left(#1\right)}
\newcommand{\IM}[1]{\text{\fontfamily{cmr}\selectfont Im}\left(#1\right)}
\newcommand{\REs}{\text{\fontfamily{cmr}\selectfont Re}}
\newcommand{\IMs}{\text{\fontfamily{cmr}\selectfont Im}}
\newcommand{\bra}[1]{\left\langle#1\vphantom{\big|}\right|}
\newcommand{\ket}[1]{\left|#1\vphantom{\big|}\right\rangle}
\newcommand{\ketbra}[2]{\left|#1\vphantom{#2\big|}\right\rangle\negthickspace\left\langle#2\vphantom{\big|#1}\right|}
\begin{document}
\begin{center}
    \textbf{One-dimensional models of disordered quantum wires: general
formalism}\\[4mm]
Alberto Rodr\'{\i}guez \footnote{email:argon@usal.es}\\
\textit{F\'{\i}sica Te\'orica. Departamento de F\'isica Fundamental. Universidad
de Salamanca. 37008 Salamanca. Spain}\\[4mm]
\begin{minipage}{0.9\textwidth}
In this work we describe, compile and generalize a set of tools that can be used to analyse the electronic
properties (distribution of states, nature of states, \ldots) of
one-dimensional disordered compositions of potentials. In particular, we derive an ensemble of universal functional equations
which characterize the thermodynamic limit of all one-dimensional models
and which only depend formally on the distributions that define the
disorder. The equations are useful to obtain relevant quantities of the system such as density of states or 
localization length in the thermodynamic limit. 
\end{minipage}%
\end{center}
\section*{Introduction}
    The pioneering work of Anderson \cite{And109_58} changed completely the
understanding of the properties of disordered systems and meant the opening
of a research field which is of primary importance nowadays.  The physics of disordered systems is currently a significant part
of condensed matter physics and it has been the subject of an intense
research activity specially during the last ten years.  Electronic
localization due to disorder is a key element to understand different physical phenomena
such as the Quantum Hall Effect or the suppression of conductivity in
amorphous matter. In the last years the effect of
the presence of statistical correlations in disordered systems has been analysed
\cite{Flo1_89,DunWu65_90,SanMac49_94,Hil30_97,IzrKro82_99,MouLyr81_98} and
the conclusions regarding the appearance of extended states in the spectrum
have been experimentally confirmed for  the
case of short-range correlations \cite{BelDie82_99} as well as for
long-range correlations \cite{KroIzr13_02,KuhIzr77_00}. Scaling theory and
Universality of the distributions of transport-related quantities
characterizing disordered systems are subjects which are still evolving
nowadays: the conditions for the validity of single parameter
scaling (SPS) have been recently reformulated
\cite{DeyEre90_03,DeyLis84_00} and it has been found that different scaling
regimes appear when  disorder is correlated \cite{DeyEre67_03}.
The presence of disorder is of key importance for the characterization 
of low-dimensional structures, such as one-dimensional quantum wires, since it plays a key role in the transport
processes and it can strongly alter the electronic properties of the system.
Unlike the case of ordered matter, for disordered systems there is a lack of a general theory
describing in a compact form their physical properties. Nevertheless a
large ensemble of different techniques exists that can be used to unravel some
features of this kind of structures. Our fundamental premise to study the electronic properties of
one-dimensional disordered systems is to consider non-interacting spinless
carriers within the independent particle approximation, that is the
Hamiltonian of the system only includes  the potential of a linear array of different
atomic units . Also our approach
focuses on the characterization of the static transport properties of
these structures. Within this framework, the aim of this work
is to describe, compile and generalize a set of tools that can be used
for all one-dimensional systems in order to analyse their electronic
properties (distribution of states, nature of states, \ldots). In
particular, we derive an ensemble of universal functional equations
which characterize the thermodynamic limit of all one-dimensional models
(within the approximations made above) and that are useful to obtain
relevant quantities of the system such as density of states or 
localization length in that limit. Therefore a great part of our efforts
are aimed at contributing to the growth of a general methodology that
can be applied  to all potential models in one-dimension.  Let us also
mention that the formalism here contained has already been used by
the author and co-workers to describe successfully a large variety of
one-dimensional disordered models
\cite{CerRod30_02,CerRod32_03,CerRod43_05,CerRod70_04,RodCer72_05}. However
a complete and general derivation of the theoretical formalism is still
lacking; the present work comes to fill this gap.

The work is organized as follows. In section \ref{sec:CTM} we make a
thorough description of the continuous transmission matrix formalism and
its applicability to finite-range as well as continuous potentials. 
Detailed analysis and calculations completing this section are contained
in appendix \ref{ap:matrix}. The canonical equation and its derivation
from the transmission matrix is treated in section \ref{sec:canonical}. In
section \ref{sec:DTM} the discrete transmission matrix formalism is briefly
commented. The reliable parameters that can be used to characterize
electronic localization are described in section \ref{sec:Localization},
where we particularly focus on the Lyapunov exponents, whose analysis is
completed in appendix \ref{ap:lyapunov}. The procedure to
calculate the distribution of states for the disordered chain is explained
and generalize in section \ref{sec:DOS} and appendix \ref{ap:DOS}, to
proceed subsequently with the construction of the functional equation
formalism which is contained in section \ref{sec:Functional} and
constitutes the main body of the work. The expressions of DOS and
 localization length in the thermodynamic limit in terms of the solutions
of the  functional equations are rigorously obtained. The applicability of
the formalism developed is illustrated studying the one-dimensional
tight-binding model.

\section{Continuous transmission matrix formalism}
\label{sec:CTM}
    The time-independent scattering process of a one-dimensional potential can be described using the
well-known continuous transfer matrix method,
\begin{equation}
    \begin{pmatrix} A_R \\ B_R \end{pmatrix}=
    \begin{pmatrix} \rmM_{11} & \rmM_{12} \\ \rmM_{21} & \rmM_{22} \end{pmatrix}
    \begin{pmatrix} A_L \\ B_L \end{pmatrix}\equiv\mathbf{M}
    \begin{pmatrix} A_L \\ B_L \end{pmatrix},    
    \label{eq:matrix}
\end{equation}
where $A_L$, $B_L$($A_R$, $B_R$), mean the amplitudes of the asymptotic
travelling plane waves $\rme^{\rmi kx}$, $\rme^{-\rmi kx}$, at the left (right) side of the
potential. The peculiarities of the transmission matrix $\mathbf{M}$ and 
its elements depend on the nature of the potential. A
detailed analysis on this subject can be found in appendix
\ref{ap:matrix}. As a summary let us say that for  real potentials 
$\mathbf{M}$ belongs to the group $\mathcal{SU}(1,1)$ and that the property
$\det\mathbf{M}=1$ holds for all kind of potentials whether they are real or complex. 

The transmission and reflection scattering amplitudes of the potential read
\begin{equation}
    t=\frac{1}{\rmM_{22}} ,\quad r^L=-\frac{\rmM_{21}}{\rmM_{22}} ,\quad
    r^R=\frac{\rmM_{12}}{\rmM_{22}},
\end{equation}
where the superscripts $L$, $R$, stand for left and right incidence.
The insensitivity of the transmission amplitude to the incidence direction 
is a universal property. In general the reflection amplitudes will differ, 
although $|r^L|=|r^R|$ for real potentials and complex ones with parity
symmetry \cite{Ahm64_01}. 

Obtaining the transmission matrix is specially easy for discontinuous
short-range potentials such as deltas or square well/barriers, for which
 the asymptotic limit is  not necessary to satisfy equation \eqref{eq:matrix}. In
these cases the effect of a composition of $N$ different potential units can be considered
through the product of their transmission matrices,
\begin{equation}
    \mathbb{M}=\mathbf{M}_N\mathbf{M}_{N-1}\cdots\mathbf{M}_2\mathbf{M}_1,
\end{equation}
therefore obtaining analytically or numerically the exact scattering probabilities
of the whole structure. This formalism can also be used to obtain the bound
states from the poles of the complex transmission amplitude.
An intuitive and general interpretation of the composition procedure can be
given in the following form. Let us consider two finite range potentials
$V_1(x)$, $V_2(x)$, characterized by the amplitudes
$t_1$, $r^L_1$, $r^R_1$, $t_2$, $r^L_2$, $r^R_2$, and 
joined at a certain point. Then, the scattering amplitudes of the composite potential can be
obtained by considering the coherent sum of all the multiple reflection processes that might
occur at the connection region \cite{Bea38_70},
\begin{subequations}
\label{eq:comp}
\begin{align}
    t &\equiv t_1\left\{\sum_{n=0}^\infty (r^L_2 r^R_1)^n\right\} t_2 =
    \frac{t_1 t_2}{1-r^L_2 r^R_1}, \\
    r^L &\equiv r^L_1 +t_1 r^L_2 \left\{\sum_{n=0}^\infty (r^L_2 r^R_1)^n\right\}
    t_1 = r^L_1 +\frac{r^L_2 t_1^2}{1-r^L_2 r^R_1}, \\
    r^R &\equiv r^R_2 + t_2 r^R_1 \left\{\sum_{n=0}^\infty (r^L_2
    r^R_1)^n\right\} t_2 = r^R_2 +\frac{r^R_1 t_2^2}{1-r^L_2 r^R_1}.
\end{align}
\end{subequations}
Replacing the scattering amplitudes with the
elements of the corresponding transmission matrices $\mathbf{M}_1$, $\mathbf{M}_2$, one can trivially 
check that in fact the latter formulae are the equations of the matrix 
product $\mathbf{M}_2\mathbf{M}_1$. Thus, the composition rules given by
\eqref{eq:comp} are not restricted to the convergence interval of the
series $\sum_{n=0}^\infty (r^L_2 r^R_1)^n$. They provide an explicit
relation of the global scattering amplitudes  in terms of
the individual former ones and can be easily used recurrently for numerical
purposes.

For continuous potentials the calculation of the transfer matrix is
more complex. After solving the Schr\"odinger equation for positive
energies, one has to take the limits $x\rightarrow\pm\infty$ to recover the
free particle states and identify the matrix elements. Hence 
equation \eqref{eq:matrix} is strictly satisfied only asymptotically. However
depending on the decay of the potential one could neglect its effects  
outside a certain length range.
\begin{figure}
    \centering
    \includegraphics[width=.55\textwidth]{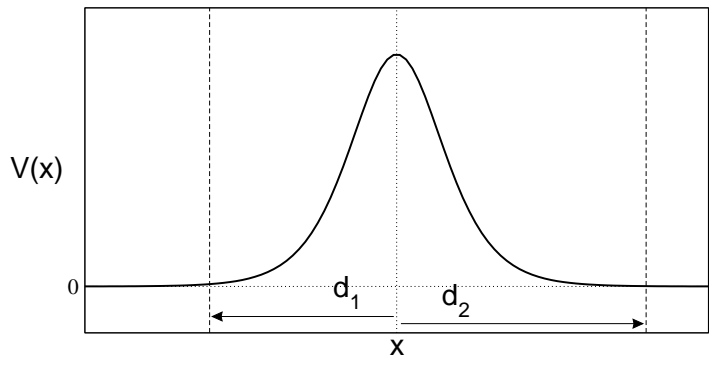}
    \caption{A continuous potential}
    \label{fig:contipot}
\end{figure}
If the asymptotic transmission matrix $\{\mathcal{M}_{ij}\}$ of the
potential in figure \ref{fig:contipot} is known, 
then the matrix for the cut-off potential contained between
the dashed lines can be written as (see appendix \ref{ap:matrix})
\begin{equation}
    \mathbf{M}_{\text{cut}}=\begin{pmatrix} \mathcal{M}_{11} \,\rme^{\rmi k(d_2+d_1)} &
                    \mathcal{M}_{12} \,\rme^{\rmi k(d_2-d_1)} \\
                    \mathcal{M}_{21} \,\rme^{-\rmi k(d_2-d_1)} &
                    \mathcal{M}_{22} \,\rme^{-\rmi k(d_2+d_1)}
                    \end{pmatrix}.
    \label{eq:matcut}
\end{equation}
The cut-off matrix is the same as the asymptotic one plus an extra phase
term in the diagonal elements that accounts for the total distance
$(d_1+d_2)$ during which the particle feels the effect of the potential,
and also an extra phase term in the off-diagonal elements measuring the
asymmetry of the cut-off $(d_2-d_1)$. Doing such approximation one gets matrices suitable to be composed in
linear arrays.
\section{The canonical equation}
\label{sec:canonical}
\begin{figure}
    \centering
    \includegraphics[width=.7\textwidth]{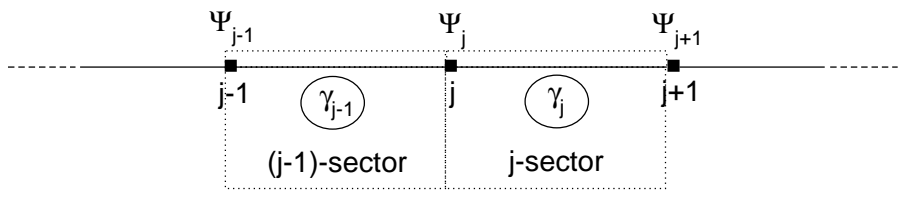}
    \caption{Sites and sectors of a linear chain}
    \label{fig:c1sectorsI}
\end{figure}

As stated in the introduction we are treating one-dimensional atomic wires within the independent particle approximation. 
The electron-electron interaction is not considered and also the carriers
are supposed to be spinless. Then, 
 the Hamiltonian of the system only includes the potential 
 of a linear array of different atomic units.
 From the solutions of the
one-particle Schr\"odinger equation it is always possible to derive an
expression with the following canonical form\footnotemark\ \cite{MACIA,SanMac49_94}
\begin{equation}
    \Psi_{j+1} = J(\gamma_{j-1},\gamma_j)\Psi_j -\frac{K(\gamma_j)}{K(\gamma_{j-1})}\Psi_{j-1},
    \label{eq:canonical}
\end{equation}
\footnotetext{The meaning of the coefficients appearing in the canonical equation 
 depends on the particular Hamiltonian. For a
tight-binding model they have  a straightforward interpretation in terms of the
on-site energies and the transfer integrals, thus the equation is
usually written in the form $\alpha_j\Psi_j = t_{j,j+1}\Psi_{j+1} +
t_{j,j-1}\Psi_{j-1}$. For other models that comparison may not be so
clear, so we keep a more general expression.}
where $\Psi_j$ means the amplitude of the electronic state at the $j$th
site of the wire, $\gamma_j$ denotes the parameters of the potential at the
$j$th site ($j$th sector) and the functions
$J(\gamma_{j-1},\gamma_j)$, $K(\gamma_j)$, which depend on the
potential and the energy, rule the spreading of the state from one site to
its neighbours, as shown in figure \ref{fig:c1sectorsI}.
The canonical equation can be systematically obtained for a given
solvable Hamiltonian  and it contains the same information as the
Schr\"odinger equation. It is not hard to see that 
$J(\gamma_{j-1},\gamma_j)$ and $K(\gamma_j)$ can be chosen to be  real functions provided the potential is real, so that the state
amplitudes can also be considered to be real. 
Equation \eqref{eq:canonical} determines also the behaviour of other 
 elementary excitations inside 1-D structures, thus it appears in
different physical contexts such as the study of vibrational states (phonons),
electron-hole pairs (excitons), \ldots

From the transmission matrix of the potential one can readily obtain the
canonical equation applying to the electronic states in the one-dimensional
composite chain. Let us consider a linear composition of potentials. All of them are
formally described by the same transmission matrix with different parameters.
And let be $\mathbf{M}_j$ the transmission matrix of the $j$th potential,
\begin{equation}
    \begin{pmatrix} A_{j+1} \\ B_{j+1} \end{pmatrix}= 
    \mathbf{M}_j \begin{pmatrix} A_j \\ B_j \end{pmatrix},
\end{equation}
 where the coordinates of the electronic wave function in the different sectors of the
chain are chosen to satisfy that the amplitude of the state at all sites is
simply given by the sum of the complex amplitudes of the travelling plane
waves, that is $\Psi_j=A_j+B_j$ for all $j$.
To build the canonical equation one simply calculates the quantity 
$\Psi_{j+1}+\chi\Psi_{j-1}$, using $\mathbf{M}_j$ and
$\mathbf{M}_{j-1}^{-1}$ to write the amplitudes $(A_{j\pm1},B_{j\pm1})$ in
terms of $(A_j,B_j)$. Then $\chi$ is solved by imposing the coefficients of
$A_j$ and $B_j$ to be the same. 
Following this procedure one concludes that the canonical equation for the
most general potential can be written as,
\begin{equation}
    \Psi_{j+1} = \left(\overline{S}_j +S_{j-1}\frac{K_j}{K_{j-1}}\right)\Psi_j -\frac{K_{j}}{K_{j-1}}\Psi_{j-1},
\end{equation}
where
\begin{subequations}
\label{eq:canomat}
\begin{align}
    \overline{S}_j &= \frac{1}{2}\left[(\mathbf{M}_j)_{11}
                +(\mathbf{M}_j)_{12}+(\mathbf{M}_j)_{21}+(\mathbf{M}_j)_{22}\right], \\
    S_j &=  \frac{1}{2}\left[(\mathbf{M}_j)_{11}
                -(\mathbf{M}_j)_{12}-(\mathbf{M}_j)_{21}+(\mathbf{M}_j)_{22}\right], \\
    K_j &=  \frac{1}{2}\left[(\mathbf{M}_j)_{11}
                -(\mathbf{M}_j)_{22}+(\mathbf{M}_j)_{21}-(\mathbf{M}_j)_{12}\right].
\end{align}
\end{subequations}
In the case of a real potential, using the symmetries of the transmission
matrix (appendix \ref{ap:matrix}), one finds
\begin{subequations}
\begin{align}
    \overline{S}_j &= \REs\left[(\mathbf{M}_j)_{11}\right] +\REs\left[(\mathbf{M}_j)_{12}\right], \\
    S_j &= \REs\left[(\mathbf{M}_j)_{11}\right] -\REs\left[(\mathbf{M}_j)_{12}\right], \\
    K_j &= \IMs\left[(\mathbf{M}_j)_{11}\right] -\IMs\left[(\mathbf{M}_j)_{12}\right].
\end{align}
\end{subequations}
And it also can be observed that for real and parity invariant potentials the
functions $\overline{S}_j$ and $S_j$ coincide because the off-diagonal
elements of the matrix are pure imaginary.
Then, the canonical equation can be easily calculated from the continuous transmission
matrix of the compositional potentials of the system.

Although the applicability of the canonical equation is not restricted by
the ordering of the sequence in the wire, it is a key ingredient to 
study non-periodic arrangements of potentials, for which the Bloch theorem
is not valid. For certain boundary conditions, one can numerically obtain the
permitted levels and the form of the envelope of the wave functions inside the system using
equation \eqref{eq:canonical}. Apart from being useful from a numerical
viewpoint, the canonical form also provides some analytical results
concerning the gaps of the system's spectrum. For this purpose, the
equation must be written as a two-dimensional mapping, originally proposed
in reference \cite{IzrKot52_95}, that permits establishing analogies between the quantum problem
and classical dynamical systems \cite{IzrRuf31_98}. The matrix form of
 \eqref{eq:canonical} with the definitions $x_{j+1}=\Psi_{j+1}$,
 $y_{j+1}=\Psi_j$, reads
\begin{equation}
    \begin{pmatrix} x_{j+1} \\ y_{j+1} \end{pmatrix}
    =\begin{pmatrix} J(\gamma_{j-1},\gamma_j) & -\frac{K(\gamma_j)}{K(\gamma_{j-1})}\\
    1 & 0 \end{pmatrix}
    \begin{pmatrix} x_j \\ y_j \end{pmatrix},
    \label{eq:map}
\end{equation}
which in polar coordinates $x_j=\rho_j\cos\theta_j$,
$y_j=\rho_j\sin\theta_j$, 
leads to the following transmission relations for the phase and the moduli:
\begin{align}
    \theta_{j+1}\equiv \Eul{T}(\theta_j;\gamma_{j-1},\gamma_j) =&   
    \arctan\left\{\left( J(\gamma_{j-1},\gamma_j)- \frac{K(\gamma_j)}
    {K(\gamma_{j-1})}\tan\theta_j\right)^{-1}\right\} \label{eq:mapfase},\\
    \left(\frac{\rho_{j+1}}{\rho_j}\right)^2 \equiv \Eul{F}(\theta_j;\gamma_{j-1},\gamma_j) =&
    \cos^2\theta_j+\left(J(\gamma_{j-1},\gamma_j)\cos\theta_j- \frac{K(\gamma_j)}{K(\gamma_{j-1})} \sin\theta_j\right)^2.
    \label{eq:maptra}
\end{align}
Now let us impose hard-wall boundary conditions in our wire composed of $N$
atoms. That means $\Psi_0=\Psi_{N+1}=0$. Using the mapping it is clear that
the initial point is $\{x_1,y_1\}=\{\Psi_1,0\}$ placed on the $x$ axis. Thus 
for an eigenenergy, after all the
steps the final point must be of the form
$\{x_{N+1},y_{N+1}\}=\{0,\Psi_N\}$ lying on the $y$ axis.
That means the whole transformation acts rotating the initial point. Therefore the permitted
levels must be clearly contained in the ranges of energy for which 
the sequence of mappings generates a rotating trajectory (generally open)
 around the origin, which is the only fixed point independently of the parameters of the mapping.
 This behaviour guarantees that after an arbitrary number of steps the final boundary
condition could still be satisfied.
However if all mappings have real eigenvalues the behaviour described is
not possible (see for example reference \cite{TABOR}). 
And it follows that permitted levels cannot lie inside the energy ranges satisfying
\begin{equation}
    J^2(\gamma_{j-1},\gamma_j) > 4 \frac{K(\gamma_j)}{K(\gamma_{j-1})}, \qquad \forall\,\gamma_j,\,\gamma_{j-1}.
    \label{eq:gaps}
\end{equation}
Note that this conclusion does not depend upon the sequence of the chain,
thus it holds for ordered and disordered structures.
\section{Discrete transmission matrix formalism}
\label{sec:DTM}
    The problem of a one-dimensional quantum wire can also be treated via a composition
procedure of another type of transfer matrices, when one obtains a
discretized version of the Schr\"odinger equation. An analytical
discretized form of this equation is given by the canonical expression 
\eqref{eq:canonical}, that can be written as
\begin{equation}
    \begin{pmatrix} \Psi_{j+1} \\ \Psi_j \end{pmatrix}
    =\begin{pmatrix}  J(\gamma_{j-1},\gamma_j)  & -\frac{K(\gamma_j)}{K(\gamma_{j-1})} \\
    1 & 0 \end{pmatrix}
    \begin{pmatrix} \Psi_j \\ \Psi_{j-1} \end{pmatrix}\equiv
    \mathbf{P}_j(\gamma_{j-1},\gamma_j) \begin{pmatrix} \Psi_j \\ \Psi_{j-1} \end{pmatrix}.
    \label{eq:canonicalDTM}
\end{equation}
The properties of the system can then be calculated from the product $\mathbf{P}_N
\mathbf{P}_{N-1}\cdots \mathbf{P}_1$ imposing appropriate boundary conditions. 

If the solutions of the differential equation are not known, one can
always take a spatial discretization, translating the original equation
\begin{equation}
     \psi^{\prime\prime}(x)=\left[V(x)-k^2\right]\psi(x),
\end{equation}
into
\begin{equation}
    \psi_{n+1}=\left\{\left[V_n-k^2\right](\Delta
    x)^2+2\right\}\psi_n-\psi_{n-1},
    \label{eq:DSE}
\end{equation}
where we have defined $\psi_n\equiv\psi(n\cdot\Delta x)$, $V_n\equiv V(n\cdot\Delta x)$   
and $\Delta x$ being the spatial step. And the corresponding matrix
representation is
\begin{equation}
    \begin{pmatrix} \psi_{n+1} \\ \psi_n \end{pmatrix}
    =\begin{pmatrix}  \left[V_n-k^2\right](\Delta
    x)^2+2 & -1\\
    1 & 0 \end{pmatrix}
    \begin{pmatrix} \psi_n \\ \psi_{n-1} \end{pmatrix}\equiv
    \mathbf{Q}_n \begin{pmatrix} \psi_n \\ \psi_{n-1} \end{pmatrix}.
    \label{eq:DSEmatrix}
\end{equation}
Then, the scattering probabilities of the system can be numerically
obtained by constructing $\mathbb{Q}=\mathbf{Q}_n \mathbf{Q}_{n-1}\cdots \mathbf{Q}_1$, considering a large enough
distance $n\cdot\Delta x$ so that the correct asymptotic form of the
state $\psi(x)=\rme^{\rmi
kx}+r \,\rme^{-\rmi kx}$ and $\psi(x)=t\,\rme^{\rmi kx}$ can be imposed at the 
extremes. The transmission and reflection probabilities are then given by
\begin{align}
    T = & \frac{4\sin^2(k\cdot\Delta x)}{\left|\mathbb{Q}_{21}-\mathbb{Q}_{12}+\mathbb{Q}_{22}
    \rme^{\rmi k\cdot\Delta x} -\mathbb{Q}_{11}\rme^{-\rmi k\cdot \Delta
    x}\right|^2},\\
    R = & \left|\frac{\mathbb{Q}_{11}-\mathbb{Q}_{22}+\mathbb{Q}_{12}
    \rme^{-\rmi k\cdot\Delta x} -\mathbb{Q}_{21}\rme^{\rmi k\cdot \Delta
    x}}{\mathbb{Q}_{21}-\mathbb{Q}_{12}+\mathbb{Q}_{22}
    \rme^{\rmi k\cdot\Delta x} -\mathbb{Q}_{11}\rme^{-\rmi k\cdot \Delta
    x}}\right|^2.
\end{align}
\section{Characterizing electronic localization}
\label{sec:Localization}

The localized nature of the electronic states inside a disordered wire can
be analyzed using different tools (see for example reference \cite{KraMac56_93}). 
Let us see some reliable parameters which can be used as a probe
of the localized or extended character of the carriers inside the
system. 

\subsection{Lyapunov exponents}
Lyapunov exponents emerge from random matrix theory \cite{CRISANTI}, and
they are used to characterize the asymptotic behaviour of systems determined by
products of such matrices. They are a key element in chaotic dynamics 
\cite{TABOR} and play an important role in the study of disordered systems.
For a full understanding of the meaning of the Lyapunov exponents and their
expressions it is mandatory to recall Oseledet's multiplicative ergodic
theorem (MET) (a complete analysis can be
found in reference \cite{ARNOLD}), which in its deterministic version and without
full mathematical rigour\footnotemark\ reads as follows. Let be $\{\mathbf{M}_n\}$ a sequence of
$d\times d$ matrices and be $\mathbb{M}_N=\mathbf{M}_N \mathbf{M}_{N-1}\cdots \mathbf{M}_1$. Then the
following matrix exists as a limit
\begin{equation}
    \lim_{N\rightarrow\infty}\left(\mathbb{M}_N^t\mathbb{M}_N\right)^\frac{1}{2N}\equiv\Gamma\geqslant 0,
\end{equation}
so that its eigenvalues can be written as
$\rme^{\lambda_1}<\rme^{\lambda_2}<\cdots<\rme^{\lambda_d}$ and the corresponding
eigenspaces $U_1,\ldots,U_d$. And for every vector $\mathbf{x}$ of this
$d$-dimensional space the following quantity exists as a limit
\begin{equation}
    \lambda(\mathbf{x})\equiv
    \lim_{N\rightarrow\infty}\frac{1}{N}\log||\mathbb{M}_N \mathbf{x}||,
\end{equation}
that verifies $\lambda(\mathbf{x})=\max(\lambda_i,\ldots,\lambda_j)$ where $\{U_i,\ldots,U_j\}$ is the set of
spaces in which $\mathbf{x}$ has a non-zero projection. 
The set $\{\lambda_i\}$ are the Lyapunov characteristic exponents
(LCE) of the asymptotic product $\mathbb{M}_N$.
\footnotetext{Several conditions must be satisfied by the set $\{\mathbf{M}_n\}$ and its
products that we suppose to be fulfilled in meaningful physical
situations.}
Therefore this theorem implies that the asymptotic exponential divergence of any spatial vector $\mathbf{x}$ 
under the action of the product of matrices $\mathbb{M}_N$ is determined by the
LCE. More precisely, the divergence will be dominated by the component of
$\mathbf{x}$ on $\{U_i\}$ with the fastest growing rate. 

Now let us consider our one-dimensional quantum wires, which as already
known can be described through products of different type of $2\times 2$
matrices, namely the discrete transfer matrices $\mathbf{P}_j$ defined from
the canonical equation in 
\eqref{eq:canonicalDTM} and the continuous
transmission matrices $\mathbf{M}_j$ defined in equation \eqref{eq:matrix}. It can
be proved that for one-dimensional Hamiltonian systems the two LCE come
in a pair of the form $\{\lambda,-\lambda\}$ (see appendix \ref{ap:lyapunov}).
Considering the discrete transfer matrices we have
\begin{equation}
    \mathbf{x}_{N+1}
    = \mathbf{P}_{N}\mathbf{P}_{N-1}\cdots\mathbf{P}_1\mathbf{x}_1,
\end{equation}
where $\mathbf{x}_{N+1}=\begin{pmatrix} \Psi_{N+1} \\ \Psi_N \end{pmatrix}$
and $\mathbf{x}_1=\begin{pmatrix} \Psi_1 \\ \Psi_0
\end{pmatrix}$. Therefore applying the MET
\begin{equation}
    \lambda(\mathbf{x_1})=\lim_{N\rightarrow\infty}\frac{1}{N}\log\sqrt{\Psi_N^2+\Psi_{N+1}^2}.
     \label{eq:lyaMET}
\end{equation}
Imposing hard-wall boundary conditions
$\lambda(\mathbf{x_1})=\lim_{N\rightarrow\infty}\frac{1}{N}\log|\Psi_N|$
which is straightforwardly equivalent to 
\begin{equation}
    \lambda =\lim_{N\rightarrow\infty}\frac{1}{N}\sum_j\log\frac{|\Psi_{j+1}|}{|\Psi_j|},
    \label{eq:lya}
\end{equation}
a common expression found in the literature for the Lyapunov exponent, and
that always provides the largest LCE \cite{TABOR}, in our case the positive one.

On the other hand the same physical system can be realized using the
continuous transmission matrix formalism, that must yield asymptotically the same values
for the Lyapunov exponents if they have physical sense at all.
Therefore,
\begin{equation}
    \mathbf{x}_{N+1}
    = \mathbf{M}_{N}\mathbf{M}_{N-1}\cdots\mathbf{M}_1\mathbf{x}_1,
\end{equation}
where  $\mathbf{x}_{N+1}=\begin{pmatrix} A_{N+1} \\ B_{N+1} \end{pmatrix}$
and $\mathbf{x}_1=\begin{pmatrix} A_1 \\ B_1\end{pmatrix}$
corresponding to the amplitudes of the travelling plane waves. If we impose
the initial conditions $A_1=1$, $B_1=r(E)$, then the final
result will be $A_{N+1}=t(E)$, $B_{N+1}=0$, where $r(E)$ and $t(E)$ are the
scattering amplitudes. Thus the MET implies
\begin{equation}
    -\lambda = \lim_{N\rightarrow\infty}\frac{1}{2N}\log T(E),
    \label{eq:lyaT}
\end{equation}
where $T(E)$ is the transmission probability of the system, and obviously
the negative Lyapunov exponent is obtained. The above expression was 
first obtained by Kirkman and Pendry \cite{KirPen17_84}. It
implies that for a given energy, the transmission of a one-dimensional disordered
structure decreases  asymptotically exponentially with the length of the system
 $T_N(E)\sim\rme^{-2\lambda(E) N}$ \cite{JohKun16_83}. This is a 
consequence of the same asymptotic exponential decreasing behaviour
exhibited by the electronic states for that energy. From this fact we define the
localization length $\xi(E)$ of the electronic state with energy $E$, if it exists, as
the inverse of the rate of the asymptotic exponential decrease of the transmission
amplitude with the length of the system for that energy,  
\begin{equation}
    \xi(E)\equiv\lambda(E)^{-1}.
    \label{eq:defloc}
\end{equation}
This definition is also a  measure of the spatial extension of the exponentially localized
state inside the system, and it has a clear physical meaning. 
Although the Lyapunov exponent  and therefore the localization length defined can only be
strictly obtained asymptotically, it makes also sense
to characterize the electronic localization in a long enough finite system through
\begin{equation}
    \xi(E)^{-1}=-\frac{1}{2N}\log T(E),
    \label{eq:lyafinite}
\end{equation}
because the Lyapunov exponent is a self-averaging quantity
\cite{KraMac56_93}, thus  it agrees with the most probable value (its mean value)
in the thermodynamic limit, for every energy. Therefore expression \eqref{eq:lyafinite} gives relevant
information of the localization length for finite $N$, since it will show a
fluctuating behaviour around the asymptotic value. 

Finally let us say that a complex extension of the Lyapunov exponent is
possible,
\begin{subequations}
\label{eq:lyacomplex}
\begin{align}
    \lambda_c &= \lim_{N\rightarrow\infty}\frac{1}{N}\sum_j\log\left(\frac{\Psi_{j+1}}{\Psi_j}\right),\\
    -\lambda_c &= \lim_{N\rightarrow\infty}\frac{1}{N} \log t(E),
    \label{eq:lyacomplexb}
\end{align} 
\end{subequations}
$\Psi_j$ being the complex amplitudes of the state and $t(E)$ the
complex transmission amplitude. The real part of this extension is related to the
localization length whereas its imaginary part turns out to be $\pi$ times the
integrated density of states $n(E)$ per length unit of the system
\cite{KirPen17_84}.
\subsection{Inverse participation ratio}
Alternatively, localization is also usually characterized by the inverse
participation ratio (IPR) \cite{CanHem18_85}, which is defined in terms of
the amplitudes of the electronic state at the different sites of the system
as
\begin{equation}
    \textrm{IPR}=\frac{\sum_{j=1}^N \left|\Psi_j\right|^4}{\left(\sum_{j=1}^N
    \left|\Psi_j\right|^2\right)^2}.
\end{equation}
For an extended state the IPR takes values of order $N^{-1}$ whereas for
a state localized in the vicinity of only one site it goes to $1$. The
inverse of the IPR means the length of the portion of the system in
which the amplitudes of the state differ appreciably from zero.
\section{Obtaining the density of states}
\label{sec:DOS}
    The density of states (DOS) $g(E)$ gives the distribution
of permitted energy levels  and it is specially important for
calculating some macroscopic properties of the structures which are usually
obtained from averages over the electronic spectrum. Strictly speaking
 $g(E)$ is the function such that $g(E)\rmd E$ is the number of eigenvalues
of the energy inside the interval $(E, E+\rmd E)$, and it is usually
defined per length unit of the system. The integrated density of states
$n(E)$ is defined as
\begin{equation}
    n(E)\equiv\int_{-\infty}^E g(E^\prime) \rmd E^\prime,
\end{equation}
and measures the number of permitted energies below the value $E$.
For a one-dimensional wire the integrated DOS can be determined from the
imaginary part of the complex Lyapunov exponent. From
\eqref{eq:lyacomplexb} one can write \cite{MACIA,SanMac49_94}
\begin{equation}
    n(E)=\frac{1}{\pi N}\arg[t^*(E)],\qquad n(E)=\frac{\rmi}{2\pi N}\log\left(\frac{t(E)}{t^*(E)}\right).
    \label{eq:idos}
\end{equation}
The electronic DOS can also be numerically determined using the negative
eigenvalue theorem proposed by Dean for the phonon spectrum \cite{Dea44_72}, however this method
cannot be applied for all potential models. It is possible to build a generalization of Dean's
method to obtain the DOS for finite chains. This technique shows some
relevant computational advantages comparing with the one involving the
complex transmission amplitude. The whole derivation can be found in
appendix \ref{ap:DOS}. Defining $s_{j,j+1}=\Psi_{j+1}/\Psi_j$ the canonical equation
\eqref{eq:canonical} reads
\begin{equation}
     s_{j,j+1} = J(\gamma_{j-1},\gamma_j) -\frac{K(\gamma_j)}{K(\gamma_{j-1})}\frac{1}{s_{j-1,j}}.
    \label{eq:ss}
\end{equation}
Now let us consider a wire composed of different atomic species $\{\alpha\}$ 
and let be $\Eul{N}_\alpha(E)$ the number of
negative $s_{j,j+1}$ whenever $\gamma_j=\alpha$, divided by the number of
sites of the chain. That is, $\Eul{N}_\alpha(E)$ is the concentration of
$\alpha$ atoms after which the envelope of the electronic wave function
with energy $E$ changes its sign.
Then the DOS per atom can be obtained as
\begin{equation}
    g(E)=\left|\sum_\alpha \sgn[K(\alpha)] \frac{\rmd
    \Eul{N}_\alpha(E)}{\rmd E}\right|.
    \label{eq:gdeE}
\end{equation}
Thus using the recursion relation \eqref{eq:ss}, one has to calculate the concentrations of changes of sing
for the different atomic species, which must then be added or subtracted according to the
sign of the functions $K(\alpha)$ for the corresponding energy. 
Finally a numerical differentiation with respect to the energy must be performed.

\section{The functional equation: the thermodynamic limit}
\label{sec:Functional}
The description of the properties of a system in the thermodynamic limit (TL) 
 reveals the fundamental physics underlying the
different problems, removing any accidental finite size effects. The TL
tells us which observations are a consequence of a general physical principle. 
With this purpose Scaling Theory is intended to obtain different
magnitudes in the TL by figuring out how they scale with the size of the
system (see reference \cite{LeeRam57_85} for a thorough description of
Scaling Theory) .
Apart from Scaling Theory, a few authors have been in pursuit
of obtaining analytically several quantities of an infinite one-dimensional
disordered system. Dyson (1953) \cite{Dys92_53} and Schmidt (1957)\cite{Sch105_57} derived
analytically a type
of functional equations for certain distribution functions 
containing information about the integrated density of states in the TL,
for the phonon spectrum of a system of harmonic oscillators with
random masses and the electronic spectrum of a delta potential model with
random couplings, respectively. Although some efforts were made to solve
numerically these equations
\cite{Aga83_64,BorBir83_64,Dea84_64,AgaBor84_64,GubTay4_71}, this approach
was almost completely forgotten probably because of its cumbersome 
mathematics and the lack of analytical solutions. 

We assert that it is possible to derive a set of universal functional
equations describing the TL of one-dimensional systems. In this way
one can build a formalism which can be applied to a large variety of potential models. The
solution of these equations can be used to obtain relevant magnitudes of
the system such as the DOS or the localization length. 
Let us begin with the canonical equation describing our one-dimensional problem,
\begin{equation}
     \Psi_{j+1} = J(\gamma_{j-1},\gamma_j)\Psi_j -\frac{K(\gamma_j)}{K(\gamma_{j-1})}\Psi_{j-1},
    \label{eq:apFEcano}
\end{equation}
where $\Psi_j$ means the real amplitude of the electronic state at the $j$th
site of the wire and $\gamma_j$ denotes the set of parameters characterizing the potential at the
$j$th site ($j$th sector). The functions $J(\gamma,\beta)$ and
$K(\gamma)$ which depend on the
potential and the energy, rule the spreading of the state from one site to
its neighbours.
From now on Greek letters are used to label the parameters of the different
types of potentials composing the chain, while Latin letters always mean
site indices. Using the mapping technique described in section
\ref{sec:canonical} one can define a phase $\theta$ and radius $\rho$
satisfying the following transmission relations:
\begin{align}
    \theta_{j+1}\equiv \Eul{T}(\theta_j;\gamma_{j-1},\gamma_j) =&   
    \arctan\left\{\left( J(\gamma_{j-1},\gamma_j)- \frac{K(\gamma_j)}{K(\gamma_{j-1})}\tan\theta_j\right)^{-1}\right\}, \\
    \left(\frac{\rho_{j+1}}{\rho_j}\right)^2 \equiv \Eul{F}(\theta_j;\gamma_{j-1},\gamma_j) =&
    \cos^2\theta_j+\left(J(\gamma_{j-1},\gamma_j)\cos\theta_j- \frac{K(\gamma_j)}{K(\gamma_{j-1})} \sin\theta_j\right)^2.
    \label{eq:apFEmaptra}
\end{align}
In order to ensure the continuity of the phase transmission for a given
energy we work with the inverse function defined as
\begin{subequations}
\begin{align}
    \Eul{T}^{-1}(\theta_{j+1};\gamma_{j-1},\gamma_j) &=
    \arctan\left\{\frac{K(\gamma_{j-1})}{K(\gamma_j)}\left(J(\gamma_{j-1},\gamma_j)-\frac{1}{\tan\theta_{j+1}}\right)\right\},\\
    \Eul{T}^{-1}(\theta_{j+1}+n\pi;\gamma_{j-1},\gamma_j) &=
    \Eul{T}^{-1}(\theta_{j+1};\gamma_{j-1},\gamma_j) \pm n\pi, \qquad
    \theta_{j+1}\in[0,\pi),\,n\in \mathbb{Z}, \label{eq:apFEtinvb}
\end{align}
\label{eq:apFEtinv}
\end{subequations}
where the plus (minus) sign in \eqref{eq:apFEtinvb} must be taken when
$\frac{K(\gamma_j)}{K(\gamma_{j-1})}>0$  $\left(\frac{K(\gamma_j)}{K(\gamma_{j-1})}<0\right)$
corresponding to an increasing (decreasing)  behaviour of the phase transmission.
\begin{figure}
    \centering
    \includegraphics[width=.8\textwidth]{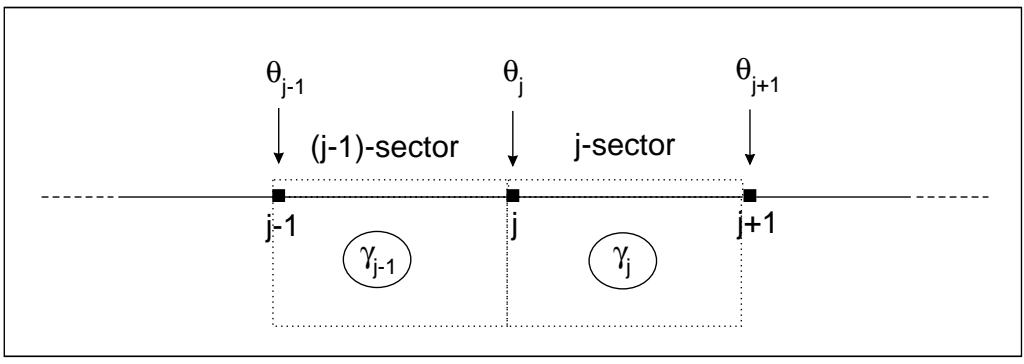}
    \caption{Sites and sectors of the one-dimensional system.}
    \label{fig:apFEsector}
\end{figure}

The goal is to calculate a distribution function for the phase $\theta$, 
valid in the thermodynamic limit, so that the differential form of such a
function acts as a natural measure of the phase in that limit. In this way
one then would be able to obtain the thermodynamic average of any quantity of the system
that could be written in terms of the phase.

The first step is to define the functions $W_j(\theta)$ with
$\theta\in[0,\pi)$, that means the probability for $\theta_j(\modu \pi)$ to
be included in the interval $[0,\theta)$, that is $\rmd W_j(\theta)$ means
the probability that $\theta_j(\modu \pi)$ belongs to $(\theta,\theta+\rmd
\theta)$, for a given energy. Therefore it follows that $W_j(\theta)$ are
monotonically increasing functions with $\theta$ such that $W_j(0)=0$, $W_j(\pi)=1$
for all $j$. And we impose
\begin{equation}
    W_j(\theta+n\pi)=W_j(\theta)+n, \qquad \theta\in[0,\pi),\,n\in\mathbb{Z},\quad \forall\, j.
    \label{eq:apFEimp}
\end{equation}
According to the meaning of these distribution functions for the individual
phases, it is clear that they must satisfy the relation
\begin{equation}
    \rmd W_{j+1}(\theta)=\rmd
    W_j\left(\Eul{T}^{-1}(\theta;\gamma_{j-1},\gamma_j)\right),
\end{equation}
that is, the probability for $\theta_{j+1}(\modu \pi)$ of being included in
$(\theta,\theta+\rmd \theta$) must be the probability for $\theta_j(\modu \pi)$ of 
appearing in $\left(\Eul{T}^{-1}(\theta), \Eul{T}^{-1}(\theta)+\rmd
\Eul{T}^{-1}(\theta)\right)(\modu \pi)$. Integrating the above equation leads us
to
\begin{equation}
    W_{j+1}(\theta)=\left|W_j\left(\Eul{T}^{-1}(\theta;\gamma_{j-1},\gamma_j)\right) - 
    W_j\left(\Eul{T}^{-1}(0;\gamma_{j-1},\gamma_j)\right)\right|,
\end{equation} 
where the absolute value is necessary for the cases when $\Eul{T}^{-1}(\theta)$ 
 decreases with $\theta$ (i.e. $[K(\gamma_j)/K(\gamma_{j-1})]<0$), because
the distribution functions must be positive. Since the inverse
transmission function of the phase gives a value in the interval
$[-\pi/2,\pi/2]$, the additional condition \eqref{eq:apFEimp} is used to
ensure that the argument of $W_j$ is always included in the interval $[0,\pi)$.
And from the definition of the inverse transmission function it follows 
\begin{equation}
    \Eul{T}^{-1}(0;\gamma_{j-1},\gamma_j)=\begin{cases}
    -\frac{\pi}{2} & \text{if } \,\frac{K(\gamma_j)}{K(\gamma_{j-1})}>0, \\
     \frac{\pi}{2} & \text{if } \,\frac{K(\gamma_j)}{K(\gamma_{j-1})}<0.
    \end{cases}
    \label{eq:apFEsigno}
\end{equation}
The equations relating the distributions for the phase at the
different sites of the system clearly show that in fact those distributions
only depend on the atomic species composing the chain. Thus the functions
can be properly redefined in terms of the compositional species. 
$W_{j+1}(\theta)$ is the distribution for the phase at the site $(j+1)$,
generated after a potential of type $\gamma_j$ (see figure
\ref{fig:apFEsector}), therefore we relabel the function as
$W_{j+1}(\theta)\equiv W_{\gamma_j}(\theta)$, that is the distribution
function for the phase after a $\gamma_j$ potential. And it is defined by
\begin{subequations}
\begin{align}
    W_{\gamma_j}(\theta) &= \left|W_{\gamma_{j-1}}\left(\Eul{T}^{-1}(\theta;\gamma_{j-1},\gamma_j)\right) 
     - W_{\gamma_{j-1}}\left(\frac{\pi}{2}\right)+\delta(\gamma_{j-1},\gamma_j)\right|, \\
    W_{\gamma_j}(\theta+n\pi) &= W_{\gamma_j}(\theta)+n, \qquad
    \theta\in[0,\pi),\,n\in\mathbb{Z},\quad \forall\,j,
\end{align}
\label{eq:apFEdefi}
\end{subequations}
where
\begin{equation}
    \delta(\gamma_{j-1},\gamma_j)=\begin{cases}
    1 & \textrm{if } \,\frac{K(\gamma_j)}{K(\gamma_{j-1})}>0, \\
    0 & \textrm{if } \,\frac{K(\gamma_j)}{K(\gamma_{j-1})}<0,
    \end{cases}
\end{equation}
due to equations \eqref{eq:apFEimp} and \eqref{eq:apFEsigno}. 
Now it is straightforward to carry out a thermodynamical average of the
probabilities $W_{\gamma_j}(\theta)$. We only have to sum over all the atomic
species and binary clusters taking into account their respective concentrations,
\begin{equation}
    \sum_{\gamma}c_\gamma
    W_\gamma(\theta)=\sum_{\gamma,\beta}C_{\beta\gamma}\left|
    W_{\beta}\left(\Eul{T}^{-1}(\theta;\beta,\gamma)\right) -
    W_{\beta}\left(\frac{\pi}{2}\right)+\delta(\beta,\gamma)\right|,
\end{equation}
where $c_\gamma$ is the concentration of the $\gamma$ species and
$C_{\gamma\beta}=C_{\beta\gamma}$ is the frequency of appearance of the
cluster --$\gamma\beta$-- or --$\beta\gamma$--. Writing
$C_{\gamma\beta}=c_\gamma p_{\gamma\beta}$, where $p_{\gamma\beta}$ is the
probability of finding a $\beta$ atom besides a $\gamma$ atom, one can
obtain an individual equation for each species,
\begin{subequations}
\begin{align}
    W_\gamma(\theta) &= \sum_{\beta}p_{\gamma\beta}\left|
    W_{\beta}\left(\Eul{T}^{-1}(\theta;\beta,\gamma)\right) -
    W_{\beta}\left(\frac{\pi}{2}\right)+\delta(\beta,\gamma)\right|, \\
    W_\gamma(\theta+n\pi) &= W_\gamma(\theta) +n, \qquad
    \theta\in[0,\pi),\,n\in\mathbb{Z}.
\end{align}
\label{eq:apFEindi}
\end{subequations}    
So that in the thermodynamic limit there exists a phase distribution function 
for each species composing the chain, and binary statistical correlations 
naturally appear in their definitions through the set of probabilities
$\{p_{\gamma\beta}\}$. The completely uncorrelated situation corresponds to
$p_{\gamma\beta}=c_\beta$ for all species. Although we have supposed a
discrete composition of the system, the same reasoning can be used for a
continuous model in which the compositional parameters belong to a certain
interval with a given probability distribution. 

Therefore solving equations \eqref{eq:apFEindi}, one would be able to
calculate the average in the thermodynamic limit of any quantity of the
system that can be written in terms of the phase $\theta$ as long as it is a periodic
function with period $\pi$. The latter expressions are the most
general functional equations valid for all
one-dimensional systems for which a canonical equation of the form
\eqref{eq:apFEcano} can be obtained.

\subsection{Calculating the localization length and the DOS in the
thermodynamic limit}
\label{sec:apFElya}
Let us consider the Lyapunov exponent given by
\begin{equation}
   \lambda = \lim_{N\rightarrow\infty}\frac{1}{N}\sum_j
    \log\left(\frac{\Psi_{j+1}}{\Psi_j}\right)\equiv \left\langle
    \log\left(\frac{\Psi_{j+1}}{\Psi_j}\right) \right\rangle,
\end{equation}
where $\langle\ \cdots\rangle$ denotes the average in the thermodynamic
limit and the amplitudes of the state at the different
sites are considered to be real. Using the two-dimensional mapping defined previously,
\begin{equation}
    \lambda = \left\langle \frac{1}{2}
    \log\left(\frac{\rho_{j+1}}{\rho_j}\right)^2\right\rangle + \left\langle
    \log\left|\frac{\cos\theta_{j+1}}{\cos\theta_j}\right|\right\rangle +
    \left\langle
    \log\left[\frac{\cos\theta_{j+1}\,|\cos\theta_j|}{|\cos\theta_{j+1}|\,\cos\theta_j}\right]\right\rangle.
\end{equation}
The middle term vanishes because the cosine is a bounded function that 
does not diverge as the length of the system grows. On the other hand the
argument of the logarithm in the last term takes only the values $\pm
1$. Since $\log(1)=0$ and $\log(-1)=\rmi \pi$ it readily follows
\begin{align}
    \RE{\lambda} &= \left\langle \frac{1}{2}
    \log\left(\frac{\rho_{j+1}}{\rho_j}\right)^2\right\rangle,\\
    \IM{\lambda} &=  -\rmi \left\langle
    \log\left[\frac{\cos\theta_{j+1}|\cos\theta_j|}{|\cos\theta_{j+1}|\cos\theta_j}\right]\right\rangle.
    \label{eq:apFEimag}
\end{align}
From equation \eqref{eq:apFEmaptra} the average of the real part can be
easily written using the distribution functions for the phase and therefore obtaining the inverse of the localization length
$\xi(E)$,
\begin{equation}
    \xi(E)^{-1}\equiv \RE{\lambda(E)} = \frac{1}{2}\sum_{\gamma,\beta}
c_\gamma p_{\gamma\beta}\int_0^\pi \rmd W_\gamma(\theta) \log
\Eul{F}(\theta;\gamma,\beta),
\end{equation}
which integrated by parts can also be written as
\begin{equation}
    \xi(E)^{-1} = \frac{1}{2} \sum_{\gamma,\beta}  c_\gamma p_{\gamma\beta}
    \log\Eul{F}(\pi;\gamma,\beta) - \frac{1}{2} \sum_{\gamma,\beta}
    c_\gamma p_{\gamma\beta} \int_0^\pi W_\gamma(\theta)  
    \frac{\Eul{F}^\prime(\theta;\gamma,\beta) }{\Eul{F}(\theta;\gamma,\beta) } \rmd \theta.
\end{equation}

On the other hand the imaginary part
of the Lyapunov exponent increases by $\rmi \pi$ every time the wave function changes sign from one site to
the next one. Therefore by averaging equation
\eqref{eq:apFEimag} over all possible species at the site $(j+1)$ when the
$j$th species is a $\gamma$ atom, and dividing by $\pi$, one obtains the fraction
of $\gamma$ atoms after which the state changes its sign.
\begin{multline}
    -\frac{\rmi}{\pi} \left\langle
    \log\left[\frac{\cos\theta_{j+1}|\cos\theta_j|}{|\cos\theta_{j+1}|\cos\theta_j}\right]\right\rangle_{j+1} \\
    = -\frac{\rmi}{\pi}\sum_\beta c_\beta \int_0^\pi \rmd W_\gamma(\theta)
    \log\left[\frac{\cos\Eul{T}(\theta;\gamma,\beta)\left|\cos\theta\right|} 
     {\left|\cos\Eul{T}(\theta;\gamma,\beta)\right|\cos\theta}\right] \\
    = \int_{\pi/2}^\pi \rmd W_\gamma(\theta) = 1-W_\gamma\left(\frac{\pi}{2}\right),  
\end{multline}
since the transmission function always returns a value in the
interval  $[-\pi/2,\pi/2]$, where the cosine is positive. Thus 
it follows that $c_\gamma \left[1-W_\gamma(\pi/2)\right]$ is
the concentration of changes of sign for the $\gamma$ species,
$\Eul{N}_\gamma(E)$ as denoted in section \ref{sec:DOS}. And from equation 
\eqref{eq:gdeE} the density of states per atom reads
\begin{equation}
    g(E)=\left|\sum_\gamma \sgn[K(\gamma)] c_\gamma \frac{\rmd
    W_\gamma\left(\frac{\pi}{2}\right)}{\rmd E}\right|.
\end{equation}
\subsection{Particular case: The canonical equation reads \mbox{$\Psi_{j+1}=J(\gamma_j)\Psi_j-\Psi_{j-1}$}}
\label{sec:apFEparti}
 The functional equations \eqref{eq:apFEindi} can be considerably simplified depending on the
particular model of the one-dimensional system. Here we consider one of the simplest forms for the canonical equation, appearing 
for example in the diagonal tight-binding model or the delta potential
model with substitutional disorder. In this case the function $J(\gamma)$ depends only on the parameters
of one potential and one can take $K(\gamma)=1$ for all 
species. Therefore the problems concerning the changes of sign of the
latter function are completely avoided and the inverse transmission
function for the phase is an increasing function for all energies which depends only on
one atomic species, $\Eul{T}^{-1}(\theta;\gamma)$. Then equations
\eqref{eq:apFEindi} read
\begin{subequations}
\begin{align}
    W_\gamma(\theta) &= \sum_{\beta}p_{\gamma\beta}\left\{
    W_{\beta}\left(\Eul{T}^{-1}(\theta;\gamma)\right) -
    W_{\beta}\left(\frac{\pi}{2}\right)\right\}+1, \\
    W_\gamma(\theta+n\pi) &= W_\gamma(\theta) +n, \qquad
    \theta\in[0,\pi),\,n\in\mathbb{Z}.
\end{align}
\end{subequations} 
If one further considers the case of uncorrelated disorder, that is
$p_{\gamma\beta}=c_\beta$ for all  $\gamma,\beta$, then a global
distribution function for the phase can be defined $W(\theta)\equiv\sum_\gamma
c_\gamma W_\gamma(\theta)$ being the solution of
\begin{subequations}
\begin{align}
    W(\theta) &= \sum_{\gamma}c_\gamma
    W\left(\Eul{T}^{-1}(\theta;\gamma)\right) -
    W\left(\frac{\pi}{2}\right)+1, \\
    W(\theta+n\pi) &= W(\theta) +n, \qquad
    \theta\in[0,\pi),\,n\in\mathbb{Z}.
\end{align}
\end{subequations} 
In this particular case only one functional equation needs to be solved and
the localization length as well as the density of states per atom can be calculated
respectively from
\begin{gather}
    \xi(E)^{-1}\equiv \RE{\lambda(E)} = \frac{1}{2}\sum_{\gamma}
    c_\gamma\int_0^\pi \rmd W(\theta) \log
    \Eul{F}(\theta;\gamma), \\
     g(E)=\left|\frac{\rmd
    W\left(\frac{\pi}{2}\right)}{\rmd E}\right|.
\end{gather}
\section{Examples}
\label{sec:tbmodel}
To exemplify the study of a quantum wire 
with the tools described in the previous sections, let us consider
a basic one-dimensional model: a tight-binding Hamiltonian with nearest neighbour 
interactions,
\begin{equation}
    \widehat{H} = \sum_k \left(\varepsilon_k \ketbra{k}{k} +t_{k,k+1} \ketbra{k}{k+1}
    +t_{k,k-1} \ketbra{k}{k-1}\right),
\end{equation}
where $\varepsilon_k$ are the energies of the on-site orbitals $\ket{k}$
and $t_{j,j\pm1}$ mean the transfer integrals, which we take equal to $1$ for
the sake of simplicity. The on-site energies follow a random sequence so
that this model is said to have diagonal disorder. The one-dimensional
Anderson model consists in choosing $\varepsilon_j$ 
from a finite continuous interval with a constant probability
distribution. In our case the composition includes different discrete species
$\{\varepsilon_1,\,\varepsilon_2,\,\ldots\}$ appearing with concentrations
$\{c_1,\,c_2,\,\ldots\}$.  

Since the orbitals $\left\{\ket{j}\right\}$ constitute an orthonormal basis of the
Hilbert space of the system, the eigenstates can be written as
$\ket{\Psi}=\sum_j u_j \ket{j}$. The Schr\"odinger equation is then
translated into a discrete equation for the coefficients $u_j$,
\begin{equation}
    u_{j+1}=(E-\varepsilon_j)u_j-u_{j-1},
    \label{eq:c1canotight}
\end{equation}
showing the desired canonical form of equation \eqref{eq:canonical} with
$J(\varepsilon_j)=E-\varepsilon_j$ and $K(\varepsilon_j)=1$ for all $j$.
Using the results of section \ref{sec:canonical} concerning the gaps of the
spectrum, it is straightforward to conclude that the energy values
satisfying $|E-\varepsilon_j|>2$ for all $j$, are not permitted. Therefore
the eigenvalues can only be located inside intervals  $4$ units of
energy wide centered at the different on-site energies. In fact each 
interval corresponds to the allowed band of the pure chain of each
species. The simplest mixed system is a binary chain composed of two
different species.  In this case the spectrum only depends upon the quantity
$|\varepsilon_1-\varepsilon_2|$ thus the on-site energies are usually
defined as
$\varepsilon_1=-\varepsilon$, $\varepsilon_2=\varepsilon$. When
$\varepsilon\leqslant 2$ the eigenenergies are all included in the
interval $[-2-\varepsilon,\varepsilon+2]$. That is the reason why this
model is commonly referred to as a one-band model. If $\varepsilon>2$ 
a gap appears in the range $[2-\varepsilon,\varepsilon-2]$.

From the canonical equation, the two-dimensional mapping defined in section
\ref{sec:canonical} is easily built yielding the transmission functions
\begin{align}
    \Eul{T}^{-1}(\theta;\varepsilon_j) &= \arctan\left(E-\varepsilon_j-\frac{1}{\tan\theta}\right),\\
    \Eul{F}(\theta;\varepsilon_j) &= 1-
    (E-\varepsilon_j)\sin(2\theta)+(E-\varepsilon_j)^2\cos^2\theta,
\end{align}
which only depend on one species at each step. This latter property
together with the fact that $K(\varepsilon_j)=1$ for this model, simplifies
considerably the functional equations \eqref{eq:apFEindi}, as described previously . For a chain
with uncorrelated disorder a unique distribution function for the phase can
be defined $W(\theta)$ being the
solution of
\begin{subequations}
\begin{align}
    W(\theta) &= \sum_\gamma c_\gamma
    W\left(\Eul{T}^{-1}(\theta;\varepsilon_\gamma)\right) -
    W\left(\frac{\pi}{2}\right) +1, \\
    W(\theta+n\pi) &= W(\theta) +n, \qquad
    \theta\in[0,\pi),\,n\in\mathbb{Z}.
\end{align}
\end{subequations}
Thus only one functional equation needs to be solved. And the DOS per atom as well as
the localization length can be obtained in the thermodynamic limit from
\begin{gather}
        \xi(E)^{-1} \equiv \lambda(E)= \frac{1}{2} \sum_\gamma  c_\gamma 
    \log\Eul{F}(\pi;\varepsilon_\gamma) - \frac{1}{2} \sum_\gamma
    c_\gamma \int_0^\pi W(\theta)  
    \frac{\Eul{F}^\prime(\theta;\varepsilon_\gamma)}
    {\Eul{F}(\theta;\varepsilon_\gamma) } \rmd \theta,\\
     g(E)=\left|\frac{\rmd W\left(\frac{\pi}{2}\right)}{\rmd E}\right|.
\end{gather}
   
\begin{figure}
    \centering
    \includegraphics[width=.55\textwidth]{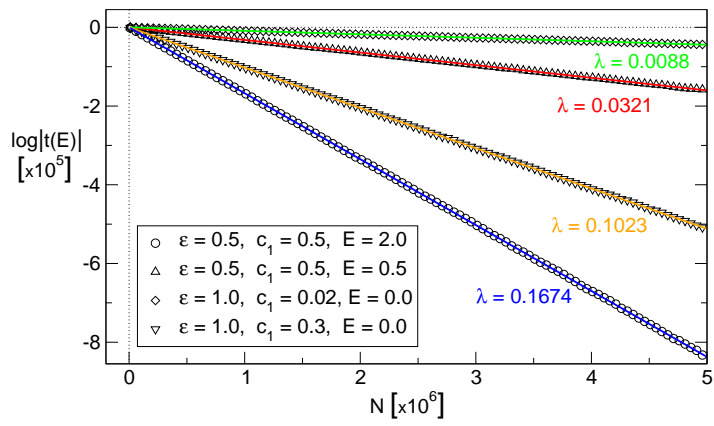}
    \caption{$\log|t(E)|$ vs length for disordered binary chains, for
    different energy values, on-site energies and concentrations. Only one
    realization of the disorder has been considered for each length. The coloured
    straight lines are plotted from the value of the Lyapunov exponent 
    in the thermodynamic limit for each case.}
    \label{fig:c1logt}
\end{figure}
\begin{figure}
\begin{minipage}{.45\textwidth}
    \centering 
    \includegraphics[width=\textwidth]{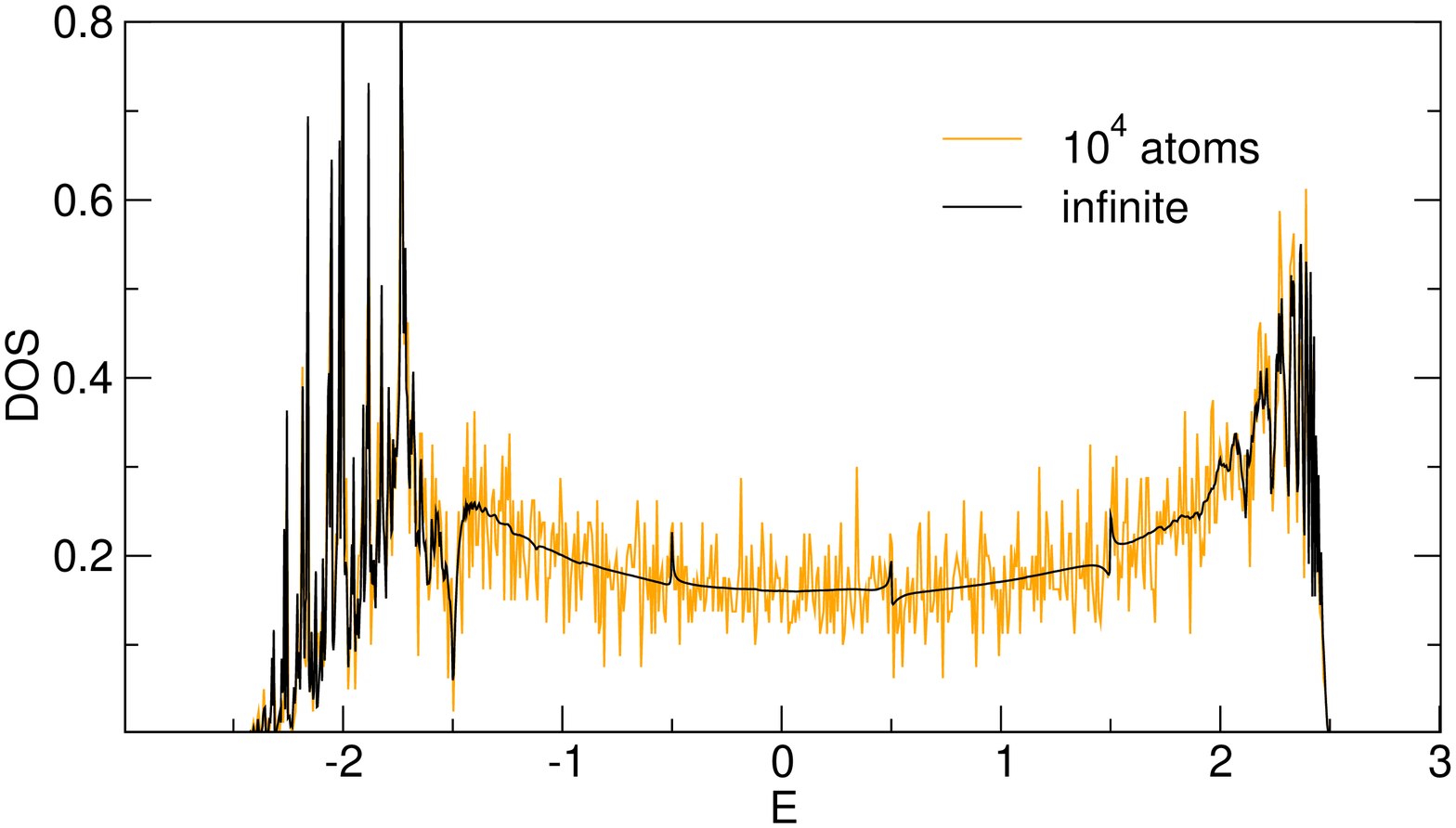}
    \caption{Comparison of the DOS for a finite binary sequence with the limiting
distribution for the infinite chain with parameters $\varepsilon=0.5,\,c_1=0.3$.}
    \label{fig:DOScompar}
\end{minipage}\hfill
\begin{minipage}{.45\textwidth}
    \vspace*{-3mm}
    \includegraphics[width=\textwidth]{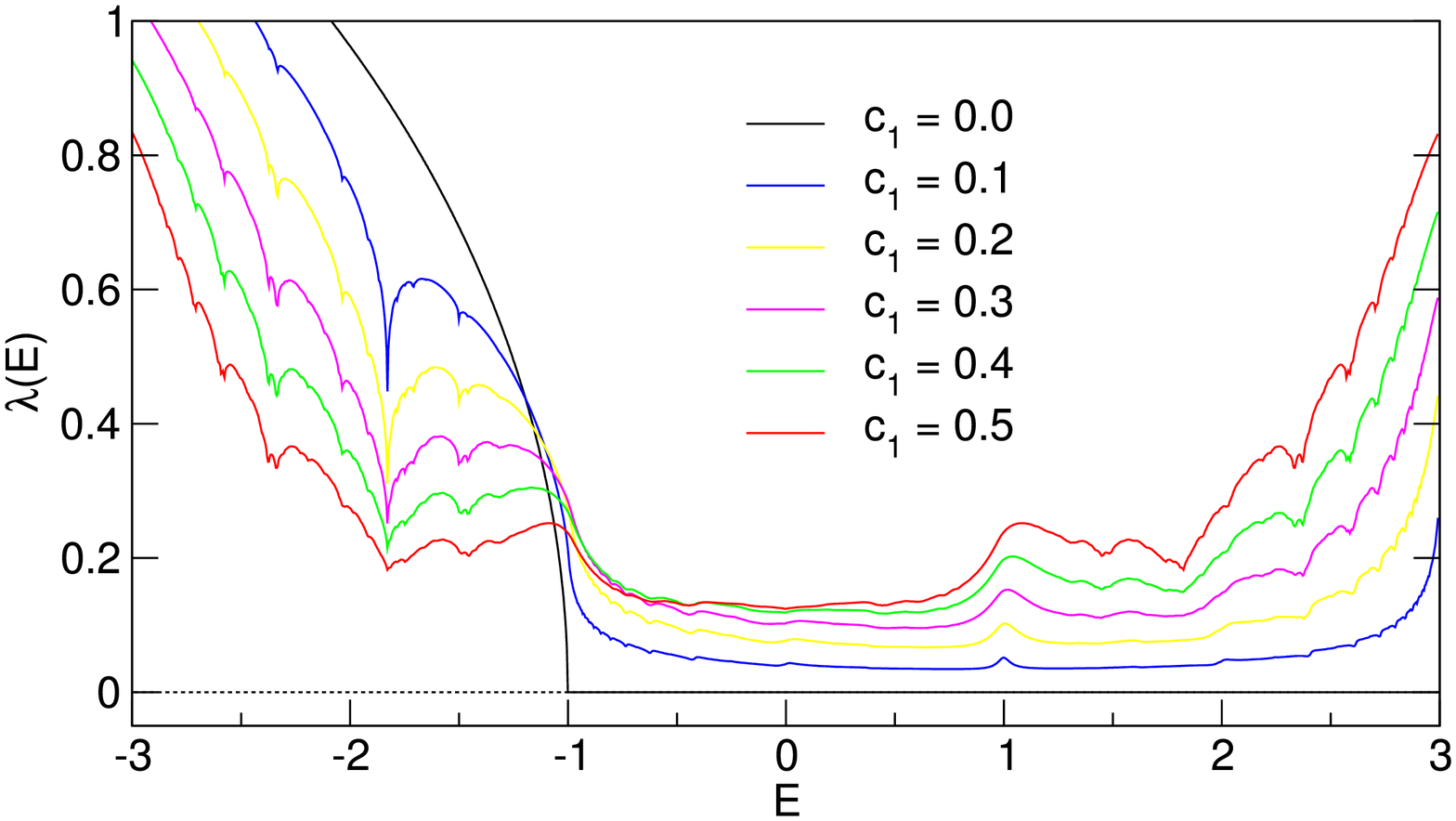}
    \caption{Lyapunov exponent for a binary chain with $\varepsilon=1$  for different
    concentrations.}
    \label{fig:c1LYAevo}
\end{minipage}
\end{figure}
\begin{figure}
    \centering
    \includegraphics[width=.6\textwidth]{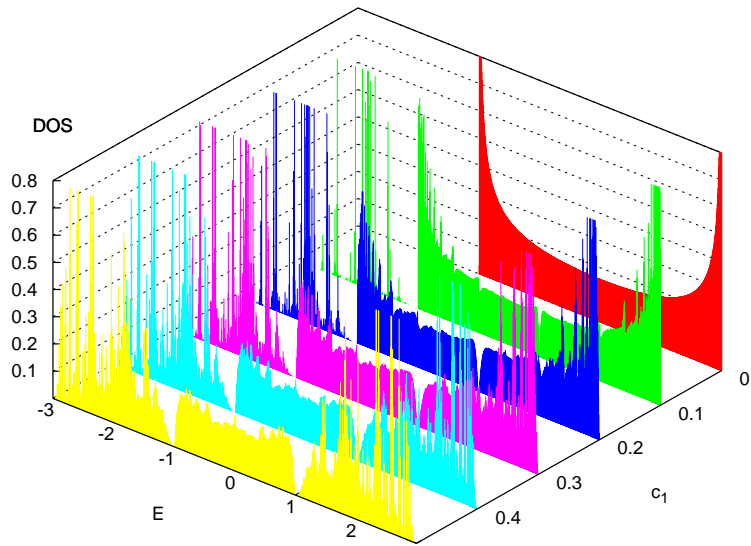}
    \caption{DOS for a binary chain with $\varepsilon=1$  for different
    concentrations. The distributions for $c_1>0.5$ are the same but inverted
    with respect to the energy due to the symmetry of the Hamiltonian.}
    \label{fig:3Dtight}
\end{figure}
 
Using the discrete transmission matrix formalism, the scattering amplitudes
of a finite chain can be obtained. In figure \ref{fig:c1logt} the logarithm
of the modulus of the transmission amplitude is plotted as a function of
the length of the system for different binary chains. The exponential
decrease of the transmission is clearly observed and the data for finite
chains agrees with the value of the Lyapunov exponent obtained from the
functional equation in the different cases. 
 In figure \ref{fig:DOScompar} the DOS for a finite chain is
 plotted, showing a fluctuating behaviour around the distribution
corresponding to the thermodynamic limit. The evolution of the DOS and the
electronic localization length in the thermodynamic limit for a
binary  chain as a function of the concentrations can be seen in figures 
\ref{fig:c1LYAevo} and \ref{fig:3Dtight}. It can be observed how the tools used provide the
correct results when the chain is composed of only one species:
$\lambda(E)=0$ and $\lambda(E)=\arc\cosh(|E-\varepsilon|/2)$ inside and
outside the allowed energy band respectively and the density of states fits the correct form
$g(E)=\pi^{-1}\left[4-(E-\varepsilon)^2\right]^{-1/2}$. 

As can be seen in the analysis of the tight-binding Hamiltonian, 
taking the canonical equation as a starting point a  systematic
characterization of the electronic
properties of the system can be performed, both for finite arrays via the
transmission matrix formalism and in the thermodynamic limit with the
functional equations. 

Aside from the tight-binding model, the functional equation formalism
described in this work has been successfully applied by the author and co-workers to other
one-dimensional potential models. It has been used to describe the DOS and
the localization of the electronic states in the thermodynamic limit of a disordered array of delta
potentials \cite{CerRod30_02}, as well as to study the significance of binary
correlations in the localization and transport properties inside the chain \cite{CerRod32_03}\cite{CerRod43_05}. 
This formalism has revealed itself as a key tool to study other non-trivial
1-D models such as the P\"oschl-Teller potential
\cite{CerRod70_04}\cite{RodCer72_05} which in the disordered
configurations exhibits a large amount of exciting properties.  We
encourage the reader to check the given references since the results there
contained are  the best proof of the usefulness and power of this methodology.
\section{Final discussion}
\label{sec:fin}
In this work we have compiled and generalized some of the existing
methodology to treat disordered quantum wires in one-dimension. The main
result  of the work is the derivation of  a set of universal functional
equations that describe analytically the thermodynamic limit of the
disordered system independently of the potential model. The 
equations only depend formally upon the distributions defining the disorder
in the system. The derivation of the functional equations has been
described in a very detailed way, starting from the transmission
matrix of the model and the canonical equation applying to the electronic
states in the system. The functional equations can be solved numerically to
carry out a systematic and thorough analysis of the electronic properties (DOS,
electronic localization, effects of short-range correlations, \ldots) of
several models of quantum wires in the thermodynamic limit,  as it has already
been done by the authors for different one-dimensional potentials
\cite{CerRod30_02,CerRod32_03,CerRod43_05,CerRod70_04,RodCer72_05}. 
The formalism developed in this work has
made it possible to observe several interesting features of the electronic
properties of one-dimensional disordered systems such as the fractal nature
of the distribution of states in the thermodynamic limit, the appearance of
extended states or the drastic change induced by  binary short-range correlations in
the DOS and the localization of the electrons within the system.  
The functional equation formalism is not restricted to the framework
of the electronic properties of disordered systems, since it can be applied
also to other Hamiltonian systems, such as for example one-dimensional spin
chains or one-dimensional phonons or excitons, which can be
described in terms of a canonical equation of the form
\eqref{eq:canonical}, where the role of the electronic state is played by a
different quantity. 

At the present time the theory of disordered systems does not include a
unifying mathematical principle comparable to the Bloch theorem for the
case of periodic structures. We would like to believe that the derivation
of the \textit{universal functional equations} for disordered systems in one
dimension may mean a little advance in this direction. The fact that
independently of the potential model it is possible to write a general set
of functional equations which only depend on the distributions defining the
disorder and that characterize the thermodynamic limit of the
systems, is to our minds a result that must be taken into account. And physical
relevant quantities such as the DOS or the localization length can be 
directly calculated in the thermodynamic limit from the functional
equations.  In spite of their formidable aspect the equations may have
analytical solutions in certain cases or they may be useful to extract 
analytically information of the systems in the thermodynamic
limit. Work along this line will probably require the use of a tough
mathematical formalism but it could also be very fruitful.

\section*{Acknowledgments}
I would like to thank my advisor J. M. Cerver\'o for several illuminating
discussions. This research has been partially supported by DGICYT under contracts
BFM2002-02609 and FIS2005-01375 and JCyL under contract SA007B05.

\appendix
\section{The transmission matrix}
\label{ap:matrix}
\subsection{Properties and symmetries of the transmission matrix}
Let be $V(x)$ a finite range potential  appreciable only inside the region
$[-d,d]$, so that the wave function can be written as
\begin{equation}
    \Psi(x)=\begin{cases}
        A_1 \rme^{\rmi k(x+d)} + B_1 \rme^{-\rmi k(x+d)}, & \quad x\leqslant -d, \\
        A_2 u(x) +B_2 v(x), & -d<x\leqslant d, \\
        A_3 \rme^{\rmi k(x-d)} + B_3 \rme^{-\rmi k(x-d)}, & \quad x> d.
        \end{cases}
\end{equation}
$u(x),v(x)$ being the linearly independent elementary solutions for each
$k$ of the continuum spectrum of the potential. 
By applying the continuity conditions of the state and its
derivative at $x=\pm d$, it is possible to reach an expression of the form
\begin{equation}
    \begin{pmatrix} A_3 \\ B_3 \end{pmatrix}=
    \begin{pmatrix} \rmM_{11} & \rmM_{12} \\ \rmM_{21} & \rmM_{22} \end{pmatrix}
    \begin{pmatrix} A_1 \\ B_1 \end{pmatrix}\equiv
    \mathbf{M} \begin{pmatrix} A_1 \\ B_1 \end{pmatrix},
    \label{eq:APmatrix}
\end{equation}
relating the amplitudes of the free particle states on the right and left
sides of the potential. $\mathbf{M}$ is the continuous transmission matrix of the
potential and its elements read in a general form:
\begin{subequations}
\label{eq:APelements}
\begin{align}
    \rmM_{11} =&
    \frac{v(d)u^\prime(-d)+v(-d)u^\prime(d)-u(d)v^\prime(-d)-u(-d)v^\prime(d)}{2\mathcal{W}}
    \notag\\ &+ \rmi\,\frac{k^2 u(d)v(-d)-k^2
    u(-d)v(d)+u^\prime(d)v^\prime(-d)-u^\prime(-d)v^\prime(d)}{2k\mathcal{W}}, \\
    \rmM_{12} =&
    \frac{v(d)u^\prime(-d)-v(-d)u^\prime(d)-u(d)v^\prime(-d)+u(-d)v^\prime(d)}{2\mathcal{W}}
    \notag\\ &+ \rmi\,\frac{k^2 u(-d)v(d)-k^2
    u(d)v(-d)+u^\prime(d)v^\prime(-d)-u^\prime(-d)v^\prime(d)}{2k\mathcal{W}}, \\
    \rmM_{21} =&
    \frac{v(d)u^\prime(-d)-v(-d)u^\prime(d)-u(d)v^\prime(-d)+u(-d)v^\prime(d)}{2\mathcal{W}}
    \notag\\ &+ \rmi\,\frac{k^2 u(d)v(-d)-k^2
    u(-d)v(d)-u^\prime(d)v^\prime(-d)+u^\prime(-d)v^\prime(d)}{2k\mathcal{W}}, \\
    \rmM_{22} =&
    \frac{v(d)u^\prime(-d)+v(-d)u^\prime(d)-u(d)v^\prime(-d)-u(-d)v^\prime(d)}{2\mathcal{W}}
    \notag\\ &+ \rmi\,\frac{k^2 u(-d)v(d)-k^2
    u(d)v(-d)-u^\prime(d)v^\prime(-d)+u^\prime(-d)v^\prime(d)}{2k\mathcal{W}},
\end{align}
\end{subequations}
where $\mathcal{W}=v(x)u^\prime(x) - u(x)
v^\prime(x)$ is the Wronskian of the solutions and it must be independent of $x$.
A straightforward calculation of the determinant of the transmission matrix
leads to
\begin{equation}
    \det\mathbf{M} =\frac{v(d)u^\prime(d) - u(d)v^\prime(d)}{v(-d)u^\prime(-d) - u(-d)
    v^\prime(-d)}.
\end{equation}
Therefore $\det\mathbf{M}=1$ for all kind of potentials, since no specific
assumptions have been made regarding $V(x)$. Let us remark that this property is not a
consequence of the time reversal symmetry of the Hamiltonian as it is
usually stated. Let us study now the
symmetries of the elements of the matrix in different special cases.
\subsubsection{Real potential ($V(x)\,\in\,\mathbb{R}$)}
If the potential is real it is always possible to find real linearly
independent solutions $u(x),v(x)$ for each value of the energy
$k$. And from equations \eqref{eq:APelements} the following relations hold,
\begin{equation}
    \rmM_{22}=\rmM_{11}^*, \qquad
    \rmM_{21}=\rmM_{12}^*.
    \label{eq:APconjugate}
\end{equation}
Therefore in the case of a real potential the transmission matrix can be
written as
\begin{equation}
    \mathbf{M}=\begin{pmatrix} \alpha & \beta \\
        \beta^* & \alpha^* \end{pmatrix} ,\quad |\alpha|^2-|\beta|^2=1.
\end{equation}
It is easy to check that these matrices satisfy
\begin{equation}
    \mathbf{M}\begin{pmatrix} 1 &0 \\0&-1\end{pmatrix}\mathbf{M}^\dagger= 
    \begin{pmatrix} 1 &0 \\0&-1\end{pmatrix}.
\end{equation}
The latter equation together with $\det\mathbf{M}=1$ define
the group $\mathcal{SU}(1,1)$.

\underline{$V(x)$ with parity symmetry}:
If the potential is real such that $V(-x)=V(x)$, then it is possible to
find elementary solutions with parity symmetry. Let us suppose $u(x)$ to be
even and $v(x)$ to be odd, then
\begin{xalignat}{2}
    u(-x) &= u(x),  & v(-x) &= -v(x), \\
    u^\prime(-x) & = -u^\prime(x), & v^\prime(-x) &= v^\prime(x).
\end{xalignat}
Using this symmetry in \eqref{eq:APelements} one finds
\begin{equation}
    \rmM_{21}=-\rmM_{12}.
    \label{eq:APparity}
\end{equation}
And together with conditions \eqref{eq:APconjugate} it yields a matrix of the form
\begin{equation}
    \mathbf{M}=\begin{pmatrix} \alpha & \rmi b \\
        -\rmi b & \alpha^* \end{pmatrix}, \quad b\,\in\,\mathbb{R},\,|\alpha|^2-b^2=1.
\end{equation}
\subsubsection{Complex potential ($V(x)\,\in\,\mathbb{C}$)}
If the potential is complex, it is not possible generally to build 
functions $u(x),v(x)$ being real, therefore the conjugation relations
\eqref{eq:APconjugate} are not satisfied. There are no special symmetries among
the matrix elements.
\begin{equation}
    \mathbf{M}=\begin{pmatrix} \alpha & \beta \\
                    \delta & \gamma 
                \end{pmatrix},\quad \alpha\gamma-\beta\delta=1.
\end{equation}
\underline{$V(x)$ with parity symmetry}:
In the case of a complex potential with parity symmetry, the same analysis
as for a real potential can be applied, and equation \eqref{eq:APparity} is
obtained, since it does not depend on the elementary solutions being real
or complex but only on their symmetries.
\begin{equation}
    \mathbf{M}=\begin{pmatrix} \alpha & \beta \\
                    -\beta & \gamma 
                \end{pmatrix},\quad \alpha\gamma+\beta^2=1.
\end{equation}
\underline{$V(x)$ with $\mathcal{PT}$-symmetry}:
Let us consider a complex local potential invariant under the joint
action of parity and time-reversal operations $V^*(-x)=V(x)$ \cite{MugPal395_04}. Then it is
possible to find $u(x),v(x)$ satisfying
\begin{xalignat}{2}
    u^*(-x) &= u(x),  & v^*(-x) &= -v(x), \\
    (u^*)^\prime(-x) & = -(u^*)^\prime(x), & (v^*)^\prime(-x) &= (v^*)^\prime(x).
\end{xalignat}
Using the above symmetries in \eqref{eq:APelements} one is led to
\begin{equation}
    \rmM_{22}=\rmM^*_{11}, \qquad
    \rmM^*_{12}=-\rmM_{12}, \qquad
    \rmM^*_{21}=-\rmM_{21}.
\end{equation}
Thus the matrix can be written as
\begin{equation}
     \mathbf{M}=\begin{pmatrix} \alpha & \rmi b \\
                    \rmi c & \alpha^* 
                \end{pmatrix},\quad b,c\,\in\mathbb{R},\,|\alpha|^2+bc=1.
\end{equation}
\subsection{Scattering amplitudes}
\begin{figure}
    \centering
    \includegraphics[width=.7\textwidth]{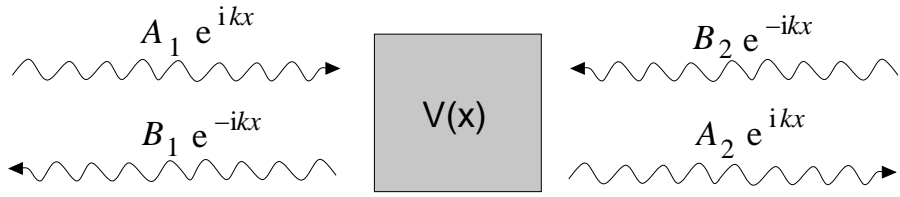}
    \caption{Representation of the scattering process.}
    \label{fig:APondas}
\end{figure}

The scattering amplitudes are directly calculated from the transmission
matrix. Figure \ref{fig:APondas} is a pictorial representation of equation
\eqref{eq:APmatrix}. 
Considering left incidence then $A_1=1$, $B_1=r^L$, $A_2=t^L$,
$B_2=0$. And it follows
\begin{equation}
    t^L=\frac{1}{\rmM_{22}}, \qquad
    r^L=-\frac{\rmM_{21}}{\rmM_{22}}.
\end{equation}   
In the case of right incidence $A_1=0$, $B_1=t^R$, $A_2=r^R$,
$B_2=1$. And the equations yield
\begin{equation}
    t^R=\frac{1}{\rmM_{22}}, \qquad
    r^R=\frac{\rmM_{12}}{\rmM_{22}}.
\end{equation}  
The insensitivity of the complex transmission amplitude to the incidence
direction is trivially proved.
Using the properties of the transmission matrix for particular cases of the
potential is easy to see that for parity invariant potentials (real or
complex) $r^L=r^R$, for generic real potentials $r^L=\rme^{\rmi\varphi}r^R$
and for generic complex ones both amplitudes differ.

In table \ref{tab:APsummary} a summary of the symmetries of the
transmission matrices and scattering amplitudes is given for the different
type of potentials described.

\begin{table}[t]
    \centering
    \label{tab:APsummary}
    \begin{tabular}{c c c}
    \hline\hline\\[-2mm]
    $\widehat{V}(x)\left[\bra{x}\widehat{V}\ket{x^\prime}=V(x)\delta(x-x^\prime)\right]$  
    & Transmission Matrix & \parbox{.2\textwidth}{\centering Scattering \\ Amplitudes}\\[2mm]\hline\\[-2mm]
    All & $\det\mathbf{M}=1$ & $t^R=t^L$\\[2mm]
      Real & $\begin{pmatrix} \alpha & \beta \\
        \beta^* & \alpha^* \end{pmatrix}\,\in\,\mathcal{SU}(1,1)$ &  $
    r^L=\rme^{\rmi\varphi}r^R$ \\[2mm]
     Real, $\left[\widehat{V},\widehat{\mathcal{P}}\right]=0$ & $\begin{pmatrix} \alpha & \rmi b \\
        -\rmi b & \alpha^* \end{pmatrix}\,\in\,\mathcal{SU}(1,1)$ & $r^L=r^R$ \\[2mm]
    Complex & $\begin{pmatrix} \alpha & \beta \\ \delta & \gamma
    \end{pmatrix}$ & $r^L\neq r^R$ \\[2mm]
    Complex, $\left[\widehat{V},\widehat{\mathcal{P}}\right]=0$  & $\begin{pmatrix}
    \alpha & \beta \\ -\beta & \gamma \end{pmatrix}$ & $r^L=r^R$\\[2mm]
    Complex, $\left[\widehat{V},\widehat{\mathcal{P}}\widehat{\mathcal{T}}\right]=0$  & $\begin{pmatrix}
    \alpha & \rmi b \\ \rmi c & \alpha^* \end{pmatrix}$ & $r^L\neq
    r^R$\\[2mm]
    \hline\hline
    \end{tabular}
    \caption{Symmetries of the transmission matrix and scattering amplitudes
    for different potentials. Greek letters mean complex
    elements of the matrix while Latin ones represent real
    coefficients. $\widehat{\mathcal{P}}$ and $\widehat{\mathcal{T}}$ denote
    the parity and time-reversal operators respectively.}
\end{table}
\subsection{Transmission matrix for a continuous potential}
For the most general continuous potential, equation \eqref{eq:APmatrix} is
only satisfied asymptotically, that is the amplitudes of the asymptotic states
\begin{subequations}
\label{eq:APinfi}
\begin{align}
    \Psi(-\infty)&=A_1\rme^{\rmi kx} +B_1\rme^{-\rmi kx},\\
    \Psi(\infty)&=A_3\rme^{\rmi kx} +B_3\rme^{-\rmi kx},
\end{align}
\end{subequations}
can be related via the
asymptotic transmission matrix $\mathcal{M}$, 
\begin{equation}
     \begin{pmatrix} A_3 \\ B_3 \end{pmatrix}=
    \mathcal{M} \begin{pmatrix} A_1 \\ B_1 \end{pmatrix}.
    \label{eq:APasym}
\end{equation}
The procedure to obtain this asymptotic matrix is the
following. First is solving the Schr\"odinger equation for positive
energies so that the more general state reads
\begin{equation}
    \Psi(x)=A_2 u(x)+B_2 v(x),
\end{equation}
in terms of the elementary solutions $u(x)$, $v(x)$. Now one needs to build the asymptotic
form of the elementary solutions,
\begin{subequations}
\begin{align}
    u(\pm\infty) &= U_1^\pm \rme^{\rmi kx} +U_2^\pm \rme^{-\rmi kx}, \\
     v(\pm\infty) &= V_1^\pm \rme^{\rmi kx} +V_2^\pm \rme^{-\rmi kx}.
\end{align}
\end{subequations}
Therefore the more general asymptotic state becomes
\begin{equation}
    \Psi(\pm\infty)=\left(A_2 U_1^\pm +B_2 V_1^\pm\right) \rme^{\rmi kx} +
                    \left(A_2 U_2^\pm +B_2 V_2^\pm\right) \rme^{-\rmi kx}.
    \label{eq:APasystate}
\end{equation} 
Equating the coefficients with those corresponding to the asymptotic
forms \eqref{eq:APinfi} yields
\begin{xalignat}{2}
    A_1 &= A_2 U_1^- +B_2 V_1^-, &  A_3 &= A_2 U_1^+ +B_2 V_1^+, \\
    B_1 &= A_2 U_2^- +B_2 V_2^-, &  B_3 &= A_2 U_2^+ +B_2 V_2^+.
\end{xalignat}
Solving $(A_3,B_3)$ in terms of $(A_1,B_1)$ one obtains for the elements of
the asymptotic matrix
\begin{equation}
    \mathcal{M}=\frac{2\rmi k}{\mathcal{W}}\begin{pmatrix} U_1^+V_2^- -V_1^+ U_2^-\, &
    V_1^+ U_1^- - U_1^+ V_1^- \\ U_2^+ V_2^- -V_2^+ U_2^-\, & V_2^+ U_1^- - U_2^+
    V_1^- \end{pmatrix},
\end{equation}
where $\mathcal{W}=v u^\prime -v^\prime u$ is the Wronskian of the solutions.
\subsubsection{Including a cut-off in the potential}
Let us suppose that due to the nature of the potential, it is only
appreciable inside the region $[-d_1,d_2]$ (figure \ref{fig:contipot}).
Then the transfer matrix for the potential with the cut-off relates the
amplitudes of the plane waves at $x=d_2$ and $x=-d_1$, which can be written
from the asymptotic forms \eqref{eq:APinfi}, yielding the relation 
\begin{equation}
    \begin{pmatrix} A_3 \rme^{\rmi kd_2} \\ B_3 \rme^{-\rmi k d_2}
    \end{pmatrix}=\mathbf{M}_{\text{cut}}
    \begin{pmatrix} A_1 \rme^{-\rmi kd_1} \\ B_1 \rme^{\rmi k d_1}\end{pmatrix},
\end{equation}
which implies 
\begin{equation}
    \mathbf{M}_{\text{cut}}=\begin{pmatrix} \rme^{\rmi k d_2} & 0 \\ 0 &
                    \rme^{-\rmi k d_2} \end{pmatrix}
    \mathcal{M}\begin{pmatrix} \rme^{\rmi k d_1} & 0 \\ 0 &
                    \rme^{-\rmi k d_1} \end{pmatrix},
\end{equation}
leading to
\begin{equation}
   \mathbf{M}_{\text{cut}}=\begin{pmatrix} \mathcal{M}_{11} \,\rme^{\rmi k(d_2+d_1)} &
                    \mathcal{M}_{12} \,\rme^{\rmi k(d_2-d_1)} \\
                    \mathcal{M}_{21} \,\rme^{-\rmi k(d_2-d_1)} &
                    \mathcal{M}_{22} \,\rme^{-\rmi k(d_2+d_1)}
                    \end{pmatrix}.
\end{equation}
Once the asymptotic transmission matrix is known, the cut-off matrix is
straightforwardly built. And for a given potential it is usually easier to calculate the asymptotic
matrix than to construct directly the cut-off version using the continuity
conditions at the cut-off points.
\section{The Lyapunov exponents}
\label{ap:lyapunov}
Let us prove that for a one-dimensional Hamiltonian system, suitable to be
described in terms of products of random matrices, the two 
Lyapunov characteristic exponents  (LCE) come in a pair of the form $\{\lambda,-\lambda\}$. 

Let us consider our system within the discrete transmission matrix
formalism. If the Hamiltonian of the system can be written in terms of a
potential $V(x)$ then it can be shown that the canonical equation takes
always the form 
\begin{equation}
    \Psi_{j+1} = J(\gamma_{j-1},\gamma_j) \Psi_j - \frac{K(\gamma_j)}{K(\gamma_{j-1})} \Psi_{j-1},
\end{equation}
where $\gamma_j$ denotes the parameters of the potential at the $j$th site
of the system and $J$, $K$ are functions depending on these parameters and  the energy.
 Hence the discrete transmission matrix reads
\begin{equation}
    \mathbf{P}_j(\gamma_{j-1},\gamma_j)=\begin{pmatrix} J(\gamma_{j-1},\gamma_j) & -\frac{K(\gamma_j)}{K(\gamma_{j-1})}\\
                    1 & 0 \end{pmatrix}. 
    \label{eq:apDTMl}
\end{equation}
The above matrix becomes a symplectic transformation if
$K(\gamma_j)/K(\gamma_{j-1})=1$ but in general $\det\mathbf{P}_j\neq 1$.
The necessary and sufficient condition to ensure that the LCE of the asymptotic
product $\mathbb{P}_N=\mathbf{P}_N\ldots \mathbf{P}_1$ ($N\rightarrow\infty$) are 
$\{\lambda,-\lambda\}$ is that the determinant of Oseledet's
matrix $\Gamma$ equals $1$. Let us remember that $\Gamma =
\lim_{N\rightarrow\infty}\left(\mathbb{P}_N^t\mathbb{P}_N\right)^\frac{1}{2N}$.
Thus we only need to prove that $\lim_{N\rightarrow\infty} (\det \mathbb{P}_N)^{1/N}=1$
which is trivial if the individual transfer matrices are symplectic, like for
example in the tight-binding model with constant transfer integrals or for
the delta potential model, but
it may not be so obvious in the most general case. For the proof let us
suppose that our system is composed of two different kind of potentials
$1$, $2$, in a random sequence. Then the product matrix $\mathbb{P}_N$ will contain four
different types of matrices, namely $\mathbf{P}(1,1)$, $\mathbf{P}(1,2)$, 
$\mathbf{P}(2,1)$, $\mathbf{P}(2,2)$ according to the different pairs of
potentials occurring in the sequence and participating in the canonical equation.
Then,
\begin{multline}
    \lim_{N\rightarrow\infty}(\det \mathbb{P}_N)^\frac{1}{N} \\
    = \lim_{N\rightarrow\infty}\left[ (\det\mathbf{P}(1,1))^{N_{11}}
    (\det\mathbf{P}(1,2))^{N_{12}} (\det\mathbf{P}(2,1))^{N_{21}}
    (\det\mathbf{P}(2,2))^{N_{22}}\right]^\frac{1}{N}\\
    = (\det\mathbf{P}(1,1))^{C_{11}}
    (\det\mathbf{P}(1,2))^{C_{12}} (\det\mathbf{P}(2,1))^{C_{21}}
    (\det\mathbf{P}(2,2))^{C_{22}},
\end{multline}
$N_{ij}$ being the number of times that $\mathbf{P}(i,j)$ appears, and
$C_{ij}$ the frequencies of appearance of the pairs $-ij-$ in the
thermodynamic limit. From \eqref{eq:apDTMl} it readily follows
\begin{equation}
    \lim_{N\rightarrow\infty}(\det \mathbb{P}_N)^\frac{1}{N}=
    \left[\frac{K(2)}{K(1)}\right]^{C_{12}}
    \left[\frac{K(1)}{K(2)}\right]^{C_{21}}.
\end{equation}
A simple reflection symmetry argument requires that
$C_{12}=C_{21}$ so the above expression equals $1$. And 
finally the eigenvalues of $\Gamma$ must be of the form
$\rme^{\lambda},\,\rme^{-\lambda}$.

This proof can be straightforwardly extended for a case considering $k$
different potentials or for a continuous model with parameters inside a certain 
range with a given probability distribution.

This result about the Lyapunov exponent for a one-dimensional system must
be naturally expected, by the fact that the same physical problem can be
treated through the continuous transmission matrix formalism and it must lead
to the same results. Therefore considering the latter matrices which have
always determinant unity, it is obvious that $\det\Gamma=1$.
\section{DOS from node counting}
\label{ap:DOS}
James and Ginzbarg obtained the expression of the integrated density of
states (IDOS) of a linear chain of potentials 
in terms of the changes of sign of the wave function inside the different 
sectors of the system \cite{JamGin57_53}. Their reasoning is the following.
Let us consider a binary wire composed of two types of potentials $\beta_1$ and
$\beta_2$. And let be $[x_j,x_{j+1}]$ the $j$th sector of the chain including
a $\beta_1$ potential. Then the elementary solutions for positive energy
in this cell $f_1(x)$, $g_1(x)$, can be chosen to satisfy
\begin{subequations}
\begin{alignat}{2}
    f_1(x_j) &=1, \qquad& f^\prime_1(x_j)&=0, \\
    g_1(x_j) &=0, \qquad& g^\prime_1(x_j) &=1.
\end{alignat}
\label{eq:apDOSele}
\end{subequations}
Now let us consider a certain energy $E$ for which $g_1(x)$ has $p_1$ nodes
inside the given sector (the first one at $x_j$). Then for this energy, 
possible states include $\psi(x)=g_1(x)+\mu f_1(x)$ with $(p_1-1)$ zeros
and $\psi(x)=g_1(x)-\mu f_1(x)$ with $p_1$ zeros in the cell (for
sufficiently small $\mu$). For low $E$, $p_1=1$ and as the energy grows the index $p_1$
increases by one whenever $g_1(x_{j+1})=0$. Thus the energy spectrum can be
divided in intervals according to the value of the index $p_1$, so that if
$E$ lies in the interval labelled with $p_1$ then the solution $\psi(x)$
will have $p_1$ or $(p_1-1)$ nodes in every sector of type $\beta_1$. 
To determine whether the number of zeros is $p_1$ or $(p_1-1)$, one has to check
if $\psi(x_j)$ and $\psi(x_{j+1})$ have the same signs (even number of nodes)  or
opposite signs (odd number of nodes). Thus for that energy the number of nodes in a
$\beta_1$ sector can be written as
\begin{equation}
    p_1-\frac{1}{2}+(-1)^{p_1}\frac{z}{2},
\end{equation}
where $z=1$ if $\psi(x_j)$ and $\psi(x_{j+1})$ have the same signs and
$z=-1$ otherwise. The same reasoning can be used for the species $\beta_2$. 
And for an energy $E$ with the labels $p_1$ for the species
$\beta_1$ and $p_2$ for $\beta_2$, the total number of nodes inside the
mixed system and therefore the IDOS per atom reads
\begin{equation}
\begin{split}
    n(E) =& \,c_1\left(p_1+\frac{(-1)^{p_1}-1}{2}\right) + 
          c_2\left(p_2+\frac{(-1)^{p_2}-1}{2}\right) \\ &-
        (-1)^{p_1} \Eul{N}_1(E) - (-1)^{p_2} \Eul{N}_2(E),
\end{split}
\end{equation}
where $c_1$, $c_2$, are the concentrations of the species and
$\Eul{N}_1(E)$, $\Eul{N}_2(E)$, are the concentrations of changes of sign for each
species, that is the number of cells containing a certain species in which
the state changes sign (i.e. $\psi(x)$ has opposite signs at the beginning
and at the end of the sector) divided by the total number of potentials of the system. To obtain the density of states one
needs to evaluate $n(E)$ in the interval $(E,E+\rmd E$) in which the indices
$p_1$, $p_2$, can be considered to remain fixed. Therefore the only quantities that
can vary in the differential interval are the concentrations of changes of sign. Hence,
\begin{equation}
    g(E)=\left|(-1)^{p_1}\frac{\rmd \Eul{N}_1(E)}{\rmd E} + 
                (-1)^{p_2}\frac{\rmd \Eul{N}_2(E)}{\rmd E}\right|.
    \label{eq:apDOSdos}
\end{equation}
And this expression is straightforwardly generalized for an arbitrary
number of species.
The main result is that to determine the DOS correctly, depending on the energy
range one has to sum or subtract
the changes of sign of the wave function at sectors corresponding to
different species.

Now let us see how one can know the indices $p_1$, $p_2$, in a practical
way. The system is completely determined by the canonical equation
\begin{equation}
   \Psi_{j+1} = J(\gamma_{j-1},\gamma_j)\Psi_j -\frac{K(\gamma_j)}{K(\gamma_{j-1})}\Psi_{j-1}.
\end{equation}
The functions $J$, $K$,  
can be obtained in terms of the elementary solutions of the Schr\"odinger
equation in each sector of the chain. Making use of equations \eqref{eq:APelements}
and \eqref{eq:canomat} the function  $K(\gamma_j)$ can be defined 
as
\begin{equation}
    K(\gamma_j)=\frac{f_{\gamma_j}(x_{j+1})g_{\gamma_j}(x_j)-f_{\gamma_j}(x_j)g_{\gamma_j}(x_{j+1})}
    {f^\prime_{\gamma_j}(x_j)g_{\gamma_j}(x_j)-f_{\gamma_j}(x_j)g^\prime_{\gamma_j}(x_j)},
\end{equation}
where $f_{\gamma_j}(x),g_{\gamma_j}(x)$ are the elementary solutions in the
$j$th sector $[x_j,x_{j+1}]$ with a potential of type $\gamma_j$.
Imposing the additional
conditions \eqref{eq:apDOSele} it follows for the case 
$\gamma_j=\beta_1$ that $K(\beta_1)=-g_1(x_{j+1})$.
That is, the function $-K(\beta_1)$ takes the same values that the elementary
solution $g_1(x)$, verifying equations \eqref{eq:apDOSele}, would reach at the end of every $\beta_1$ sector.
Thus, whenever $g_1(x_{j+1})$ as a function of the energy changes its sign, and therefore
the index $p_1$ increases by $1$, $K(\beta_1)$ also registers a change of
sign. What is more, it is not hard to see that in equation
\eqref{eq:apDOSdos} the terms $(-1)^{p_1}$, $(-1)^{p_2}$, can be directly
identified with the signs of $K(\beta_1),\,K(\beta_2)$. And finally, one
can write
\begin{equation}
    g(E)=\left|\sgn[K(\beta_1)]\frac{\rmd \Eul{N}_1(E)}{\rmd E} + 
                \sgn[K(\beta_2)]\frac{\rmd \Eul{N}_2(E)}{\rmd E}\right|.
\end{equation}

From a numerical viewpoint one must do the transmission of the state
through the system using the functional equation and count the number of
changes of sign from site to site for the different atomic
species, to perform finally a numerical differentiation with respect to the
energy and sum or subtract the different contributions of the species
according to the sign of the function $K(\beta)$ for the energy considered.

\end{document}